\documentclass[12pt]{spieman}  
\usepackage{amsmath,amsfonts,amssymb}
\usepackage{graphicx}
\usepackage{setspace}
\usepackage{tocloft}

\newcommand{\wrt}[1]{\mathrm{d}{#1}}
\newcommand{\power}{GW/cm$^2$}
\usepackage[version=4]{mhchem}

\title{Photonics of Time-Varying Media}

\author[a,b,*]{Emanuele Galiffi}
\author[b]{Romain Tirole$^\dagger$}
\author[a]{Shixiong Yin$^\dagger$}
\author[a,c]{Huanan Li}
\author[b]{Stefano Vezzoli}
\author[d]{Paloma A. Huidobro}
\author[d]{M\'ario G. Silveirinha}
\author[b]{Riccardo Sapienza}
\author[a,e]{Andrea Al{\`u}}
\author[b]{J. B. Pendry}
\affil[a]{Photonics Initiative, Advanced Science Research Center, City University of New York, 85 St. Nicholas Terrace, 10031, New York, NY, USA}
\affil[b]{The Blackett Laboratory, Department of Physics, Imperial College London, Prince Consort Road SW2AZ, London, UK}
\affil[c]{School of Physics, Nankai University, Tianjin 300071, China}
\affil[d]{Instituto de Telecomunica\c c\~oes, Instituto Superior Tecnico-University of Lisbon, Avenida Rovisco Pais 1, Lisboa, 1049-001 Portugal}
\affil[e]{Physics Program, Graduate Center,  of the City University of New York, New York, NY 10016, USA}

\cftpagenumbersoff{figure}
\cftpagenumbersoff{table} 
\begin{document} 
\maketitle

\begin{abstract}
Time-varying media have recently emerged as a new paradigm for wave manipulation, thanks to the synergy between the discovery of novel, highly nonlinear materials, such as epsilon-near-zero materials, and the quest for novel wave applications, such as magnet-free nonreciprocity, multi-mode light shaping, and ultrafast switching. In this review we provide a comprehensive discussion of the recent progress achieved with photonic metamaterials whose properties stem from their modulation in time. We review the basic concepts underpinning temporal switching and its relation with spatial scattering, and deploy the resulting insight to review photonic time-crystals and their emergent research avenues such as topological and non-Hermitian physics. We then extend our discussion to account for spatiotemporal modulation and its applications to nonreciprocity, synthetic motion, giant anisotropy, amplification and other effects. Finally, we conclude with a review of the most attractive experimental avenues recently demonstrated, and provide a few perspectives on emerging trends for future implementations of time-modulation in photonics. 
\end{abstract}


{\noindent \footnotesize\textbf{*}Emanuele Galiffi, \linkable{egaliffi@gc.cuny.edu}.\\ }
{\noindent \footnotesize\textbf{$\dagger$} These authors contributed equally.}

\tableofcontents 

\begin{spacing}{2}   

\section{Introduction}
\label{sect:intro}  

In the wake of the extraordinary scientific advances of the last two centuries, the role of time underpins many of the unsolved puzzles of the physical world. In quantum theory, time is the only quantity not associated to an observable, being merely treated as a variable, while attempts to bring it to a common ground with all other physical quantities are still in the making~\cite{knox2021routledge}. Meanwhile, at cosmological scales, the measured acceleration of the expansion of the universe makes it hard to imagine a complete cosmological theory where temporal dynamics of e.g. wave phenomena does not play a central role. Yet, most basic physics is carried out under the assumption that physical systems are passive, merely responding ``after" any input stimuli.

Occasionally, the need to account for certain truly multi-physics or nonlinear phenomena brings up opportunities to develop our understanding of wave physics in temporally inhomogeneous systems: for instance, in order to realize bench-top analogues of relativistic phenomena early attempts were made to tap into time-modulated classical wave systems~\cite{schutzold2005hawking,philbin2008fiber}. However, the critical mass of research efforts needed to accomplish sizeable advances in this direction has, until recently, been lacking. On yet another front of exploration, time-crystals have been proposed in condensed matter theory for almost a decade as a stable state of matter capable of spontaneously breaking continuous time-translation symmetry while preserving long-range temporal order~\cite{shapere2012classical}: their very existence has been questioned and revised~\cite{sacha2017time}, and several increasingly successful implementations have recently been accomplished~\cite{yao2020classical,kessler2021observation}.

Meanwhile, over the past two decades, the rise of the field of metamaterials has proven how much fundamental wave physics lies untapped under the hood of well-established classical theories such as electromagnetism, acoustics and elasticity. Recently, the quest for novel forms of wave-matter interactions in these fields has led to a significant growth of interest in the exploration of time as a novel degree of freedom for metamaterials~\cite{engheta2021metamaterials}. In this pursuit, our need for even a basic understanding of wave phenomena in the time domain has reflected in the wealth of surprising physical effects recently unveiled, largely theoretically and in part experimentally, which can be enabled by externally applying explicit temporal modulations on the parameters of a physical system. 

In this Review Article we present an overview of the state of the art across the rapidly expanding field of time-varying photonic metamaterials, in the hope that it may serve the photonics and metamaterials community as a catalyst for current and future explorations of this fascinating field. Whilst not as detailed as a tutorial paper, the review aims at presenting the basic building blocks of time-varying electromagnetics, with the main goal of developing basic insight into the relevant phenomenology, while offering a broad view of key past and future research directions in the field. 

The paper is structured as follows: Section \ref{sec:switching} starts from the basics of electromagnetic time-switching, expanding towards more complex scenarios such as such as non-Hermitian and anisotropic switching, temporal slabs and space-time interfaces. Section \ref{sec:timeperiodic} is dedicated to photonic time-crystals, namely electromagnetic systems undergoing infinite (periodic or disordered) modulations of their parameters in time-only, and we discuss some of its most promising directions towards e.g. the engineering of topological phases, synthetic frequency dimensions, non-Hermitian physics and localization, concluding with some perspective on the peculiarities of time-varying surfaces. Spatiotemporal crystals are discussed in Sec. \ref{sec:spacetime}, where we introduce the peculiarities of space-time band diagrams, homogenization procedures for travelling-wave modulated systems enabling effective-medium descriptions of synthetic motion, and highlight some recently emerged opportunities for nonreciprocity, synthetic Fresnel drag and wave amplification and compression, as well as space-time modulated metasurface and some of their applications. Finally, Sec. \ref{sec:experiments} offers an overview of the most promising platforms for experimental implementations of time-modulation in photonics.

\newpage

\section{Wave engineering with temporal interfaces}
\label{sec:switching}

\begin{figure}
    \centering
    \includegraphics[width=0.6\textwidth]{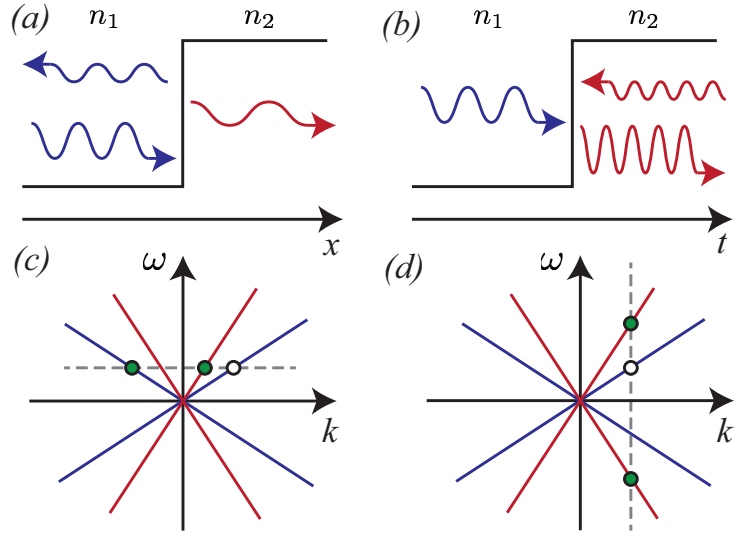}
    \caption{Scattering at (a) a spatial discontinuity, generating transmitted and reflected waves propagating in different media with refractive indices $n_1$ and $n_2$, and (b) a temporal discontinuity, generating forward and backward waves propagating in the same medium. (c,d) Illustration of the respective scattering processes on a dispersion diagram: the blue and red lines denote the dispersion cone for the two media. Note how the frequency $\omega$ is conserved in a spatial scattering process, and coupling occurs to forward and backward modes in the two different media, whereas in a time-scattering process the momentum $k$ is conserved, and the two scattered waves are both embedded in the new medium. }
    \label{fig:introFigure}
\end{figure}

Light scattering by spatial interfaces [Fig.~\ref{fig:introFigure}(a)] is at the basis of many wave phenomena, from the simplest refraction effects to the complexity of wave engineering in complex media involving multilayers\cite{novotnyPrinciples2012}, anisotropy\cite{salehFundamentals2019}, nonlinearity\cite{boydNonlinear2020}, extreme constitutive parameters\cite{veselagoElectrodynamics1967,pendryNegative2000,aluEpsilon2007}, or metamaterials and metasurfaces\cite{iyerMetamaterials2020}, to name a few. How do electromagnetic waves behave when encountering interfaces in time rather than in space? Temporal interfaces can be created by abruptly switching the material properties of a medium in time, preserving its spatial continuity [Fig.~\ref{fig:introFigure}(b)]. Wave propagation in switched media was firstly investigated by Morgenthaler in 1958\cite{morgenthalerVelocity1958}. In parallel, the study of light interactions with abruptly switched materials was developed in the plasma physics community, since the plasma permittivity can be switched in time by fast ionization processes\cite{mendonccaTheory2000}. In this section, we focus on the recent explorations on time-switching in photonic materials.

\subsection{Temporal boundary conditions and temporal scattering}

It is well-known that wave propagation across spatial interfaces is governed by established boundary conditions, see e.g., Ref.~\citeonline{jacksonClassical2012}. The essential constraint is that all relevant physical quantities, for instance, electric and magnetic fields ($\mathbf{E}$ and $\mathbf{H}$), must be finite at all points in space and time. Spatial boundary conditions are therefore derived by integrating Maxwell’s equations over an infinitesimal area and volume across the spatial interface, implying the continuity of tangential $\mathbf{E}$ and $\mathbf{H}$ at an interface that does not support surface current densities. Dually, the integration over a time interval across a temporal boundary obeys temporal boundary conditions. Let us start with the macroscopic Maxwell’s curl equations:
\begin{align}
    \frac{\partial \mathbf{B}}{\partial t}  &= -\nabla \times \mathbf{E},
    \label{eq:Faraday} \\
    \frac{\partial \mathbf{D}}{\partial t} &= -\mathbf{J}+\nabla \times \mathbf{H}.
    \label{eq:Ampere}
\end{align}
\noindent
Here we assume that the medium is unbounded and homogeneous before and after a switching instance at $t=t_0$. Integrating from $ t_0^-=t_0-\epsilon $ to $ t_0^+=t_0+\epsilon $ with a vanishing $\epsilon$, we expect that the right-hand sides of Eqs. \ref{eq:Faraday} and \ref{eq:Ampere} are both zero, due to the finite values of fields and sources. Then we obtain the temporal boundary conditions:
\begin{subequations}
\begin{align}
    \mathbf{B} \left( t=t_0^+ \right) & = \mathbf{B} \left( t=t_0^- \right), \\
    \mathbf{D} \left( t=t_0^+ \right) & = \mathbf{D} \left( t=t_0^- \right).
\end{align}
\label{eq:time_BC}
\end{subequations}
\noindent
They ensure that the magnetic flux density $\mathbf{B}$ and the electric displacement $\mathbf{D}$ vary continuously in the time domain. Alternatively, the continuity of $\mathbf{B}$ and $\mathbf{D}$ can also be interpreted by the conservation of magnetic flux $\Phi$ and electric charges $Q$, as shown by Morgenthaler\cite{morgenthalerVelocity1958} and Auld\cite{auldSignal1968}, respectively.

One of the most intriguing phenomena that the temporal discontinuity in a medium brings about is that it can scatter waves in a way dual to spatial interfaces, producing forward and backward waves in time, rather than waves reflected and transmitted in space\cite{morgenthalerVelocity1958,auldSignal1968,fanteTransmission1971,fanteOn1973,mukherjeeReflection1973,xiaoReflection2014}. The scattering coefficients can be found by applying the above temporal boundary conditions. Consider a monochromatic plane wave traveling in an unbounded, homogeneous, isotropic, non-dispersive but time-varying medium. Both permittivity $\varepsilon$ and permeability $\mu$ are assumed to abruptly switch from their initial values $\varepsilon_1$ and $\mu_1$ to $\varepsilon_2$ and $\mu_2$ at $t=t_0$. We can denote the incident wave with $D_x=D_1 e^{j\omega_1 t-jkz}$ and $B_y=B_1 e^{j\omega_1 t-jkz}$, where $\omega_1$ is the angular frequency and $k$ is the wavenumber. The field amplitudes are related by $B_1=Z_1 D_1$ with $Z_1=\sqrt{\mu_1/\varepsilon_1}$. The fields after the switching event read
\begin{subequations}
\begin{align}
    D_x \left( t>t_0 \right) & = \left[ T e^{j\omega_2 \left(t-t_0\right)} + R e^{-j\omega_2 \left(t-t_0\right)} \right] D_1 e^{j
        \left(\omega_1 t_0 - k_2 z \right)}, \\
    B_y \left( t>t_0 \right) & = Z_2\left[ T e^{j\omega_2 \left(t-t_0\right)} - R e^{-j\omega_2 \left(t-t_0\right)} \right] D_1 e^{j\left(\omega_1 t_0 - k_2 z \right)}.
\end{align}
\label{eq:DBfields}
\end{subequations}
$T$ and $R$ are the transmission and reflection coefficients defined for the electric displacement field, and $Z_2=\sqrt{\mu_2/\varepsilon_2}$ is the new wave impedance after switching. The temporal boundary conditions (Eq. \ref{eq:time_BC}) need to be satisfied everywhere in space, which implies that $k_2=k$, or equivalently
\begin{align}
    \omega_2 \sqrt{\varepsilon_2 \mu_2} &= \omega_1 \sqrt{\varepsilon_1 \mu_1}.
    \label{eq:mometum_consv}
\end{align}
This condition ensures momentum conservation across temporal interfaces, which is expected due to the preserved spatial homogeneity\cite{morgenthalerVelocity1958,mendonccaTheory2000,salemSpace2015,calozSpacetime2020a}. By equating the incident and scattered $\mathbf{D}$ and $\mathbf{B}$ fields at $t=t_0$, we solve for the temporal scattering coefficients:
\begin{subequations}
\begin{align}
    T & = \frac{Z_2 + Z_1}{2Z_2},\\
    R & = \frac{Z_2 - Z_1}{2Z_2}.
\end{align}
\label{eq:Coeff}
\end{subequations}
\noindent
The coefficients for the electric field can be easily found by substituting the constitutive relation between $D$ and $E$, which are adopted more widely in the literature. It should be emphasized that the results presented in Eqs. \ref{eq:mometum_consv} and \ref{eq:Coeff} generally apply to an abrupt temporal interface under the assumptions of spatial homogeneity, isotropy and instantaneous response of materials, and may change if one or more of these assumptions are relaxed\cite{bakunovReflection2014}. Illustrations of spatial and temporal scattering processes are given in Fig.~\ref{fig:introFigure}(a,c) and (b,d) respectively. As we consider more realistic and complicated electrodynamic models of materials, such as anisotropy, dispersion, and broken translational symmetry in space, the temporal scattering is expectedly modified\cite{kalluriElectromagnetics2018,akbarzadehInverse2018,bakunovLight2021,SolisTime2021,gratusTemporal2021,zhangTemporal2021,zhangTime2021}. To date, research on these topics is very active, as discussed in the following subsections.

\subsection{Temporal slabs}

Leveraging scattering phenomena at time interfaces, research efforts have been dedicated to the engineering of wave interference in time, for example in the context of discrete time crystals\cite{elseFloquet2016,yaoDiscrete2017} and time metamaterials\cite{pacheco2021temporal}. Even when multiple temporal interfaces are involved, temporal interferences occur only between forward and backward waves induced by each boundary independently, due to the irreversibility of time, which introduces significant differences compared to spatial interfaces. A key contrast between spatial and temporal scattering from a mathematical standpoint is that space-scattering generally gives rise to boundary-value problems, whereas time-scattering leads to initial-value problems. Here we discuss the recent progress in temporal slabs comprising one or two consecutive switching events. In Sec. \ref{sec:timeperiodic}, we discuss (quasi-)periodically switched media.

\begin{figure}[t]
    \centering
    \includegraphics[width=\textwidth]{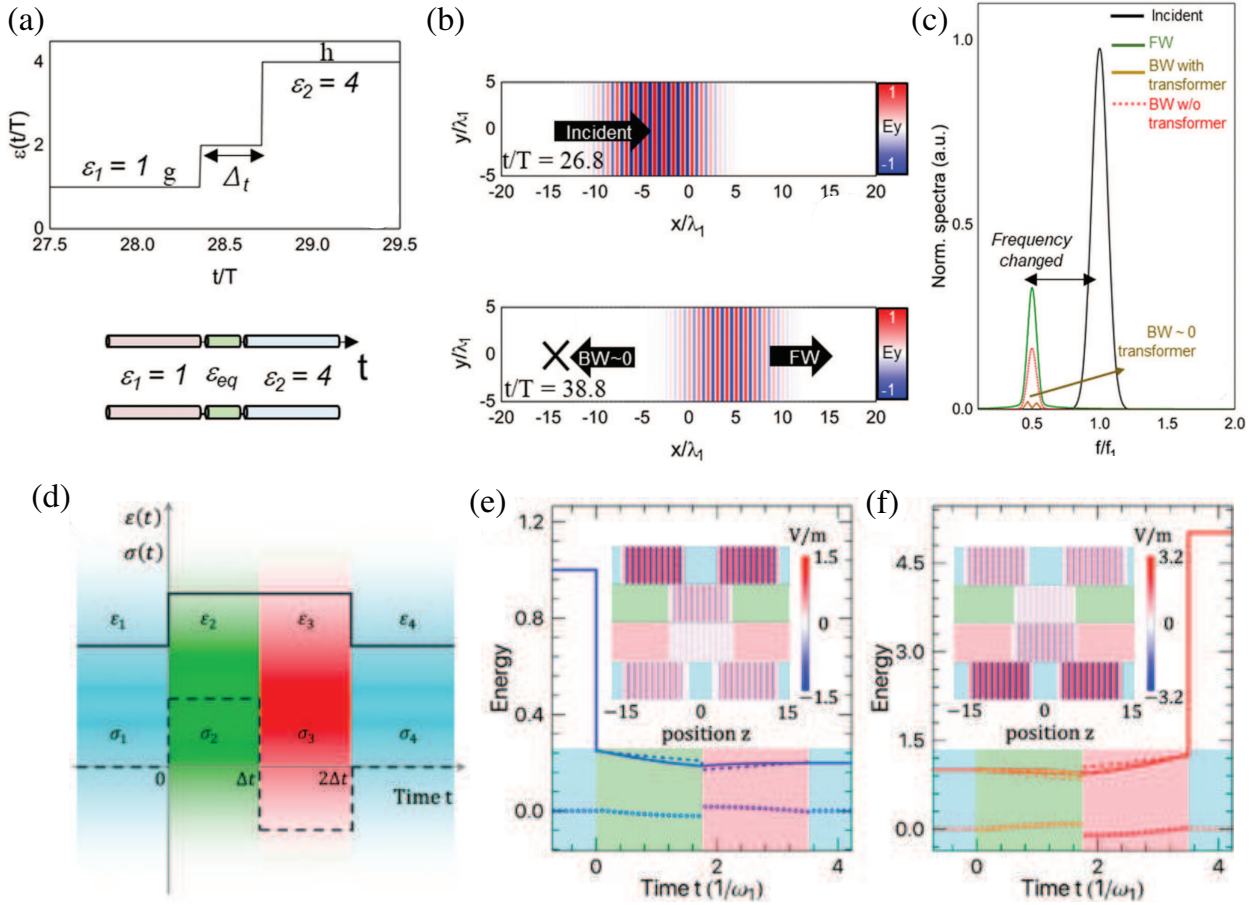}
    \caption{(a) Schematic of antireflection temporal coatings proposed in Ref.~\citeonline{pachecoAntireflection2020}. (b) Numerical results for the field distribution before and after the temporal scattering, showing the incident and transmitted waves, with minimized reflection due to the temporal coating shown in (a). (c) Spectra of the incident, forward and backward waves. (d) Schematic of temporal parity-time (TPT) symmetric structures proposed in Ref.~\citeonline{liTemporal2021}. (e) and (f): The time evolution of the normalized energy (solid curves) and its two constitutions (circle symbols for interferences and the dashed curves for another). (e) depicts the case of maximum total power while (f) shows that of minimum total power. Insets are the field distributions in simulations before, during and after the TPT slabs. Figures adapted from Refs.~\citeonline{pachecoAntireflection2020,liTemporal2021}.}
    \label{fig:temporalSlab}
\end{figure}

The temporal interference introduced by a single temporal slab was firstly investigated by Mendon{\c{c}}a and Martins in Ref.~\citeonline{mendonccaTemporal2003}, where they switched the refractive index of a medium from $n_0$ to $n_1$, and then switched it back to $n_0$ after a time interval $\tau$. The total transmission and reflection amplitudes were found to be
\begin{subequations}
\begin{align}
    T_{total} & = \left[ \cos\left(\omega_1\tau\right) + \frac{j}{2\alpha}
        \left( 1+\alpha^2 \right)\sin\left(\omega_1\tau\right) \right]e^{-j\omega_0\tau}, \\
    R_{total} & = - \frac{j}{2\alpha}
        \left( 1+\alpha^2 \right)\sin\left(\omega_1\tau\right)e^{j\omega_0\tau},
\end{align}
\label{eq:SlabCoeff}
\end{subequations}
\noindent
where $\alpha=n_0/n_1$ and $\omega_0$, $\omega_1$ are the frequencies of waves outside and inside of the temporal slab, respectively. Similar results were also obtained using a quantum optics formulation, illustrating the probability of creating Fock states of photon pairs with opposite momentum from the vacuum state\cite{mendonccaTemporal2003}. A temporal Fabry-Perot slab with dispersion was recently studied in Ref.~\citeonline{zhangTime2021}, with slightly different results due to dispersion. The most important phenomenon induced in these systems may arguably be the amplification of light. Because frequency and energy are not conserved at time boundaries, we can indeed expect large amplification of the input energy at a suitably tailored temporal slab. It can be seen from Eq. \ref{eq:SlabCoeff} that such amplification process attains the maximum level when time interval $\omega_1\tau$ is an odd multiple of $\pi/2$. On the contrary, time reflections can be minimized with a different choice of switching intervals. This finding can be connected to the dual phenomenon of impedance-matching layers in spatial scattering. Indeed, Pacheco-Pe{\~n}a and Engheta have recently introduced antireflection temporal coatings by rapidly changing the permittivity of an unbounded medium twice\cite{pachecoAntireflection2020}, making an analogy to the quarter-wavelength impedance matching in the spatial domain. The schematic of such coatings is shown in Fig.~\ref{fig:temporalSlab}(a). To minimize the backward reflection when the permittivity is to be switched from $\varepsilon_1$ to $\varepsilon_2$, they introduced a temporal coating with permittivity $\varepsilon_{eq}$ and duration $\Delta t$ in between. Impedance matching is enabled with
\begin{align}
    \varepsilon_{eq} &= \sqrt{\varepsilon_1 \varepsilon_2}, \Delta t = \frac{nT_{eq}}{4}; n=1,3,5,...
    \label{eq:coating}
\end{align}
where $T_{eq}=1/f_{eq}$ is the equivalent period of the wave right after the permittivity is switched to $\varepsilon_{eq}$, and the equivalent frequency $f_{eq}=\sqrt{\varepsilon_1/\varepsilon_{eq}} f_1$. A numerical simulation is shown in Fig.~\ref{fig:temporalSlab}(b), indicating nearly vanishing reflection from the temporal interfaces. Although Eq. \ref{eq:coating} looks similar to the condition for impedance matching in the spatial case\cite{pozarMicrowave2011}, it is more than a dual to its spatial counterpart. Firstly, the frequency is shifted from $f_1$ to $f_2=\sqrt{\varepsilon_1/\varepsilon_2} f_1$, as shown in Fig.~\ref{fig:temporalSlab}(c). In addition, the energy and power flow of the waves have also been modified as we change the permittivity of the material.

More recently, it has been shown that extreme power and energy manipulation can be achieved in these types of temporal slabs once loss and gain are considered. In analogy to conventional parity-time (PT) symmetry with balanced gain and loss\cite{christodoulidesParity2018}, Li \textit{et al.} investigated a scenario where a pair of temporal slabs obey temporal parity-time (TPT) symmetry\cite{liTemporal2021}. In their work, the temporal parity operation flips the arrow of time, while the “time” operation is defined to reverse the waves in space. TPT-symmetric temporal slabs are therefore realized if an unbounded medium is switched from a positive to a negative conductivity in time, $\sigma_2=-\sigma_3>0$, for equal duration $\Delta t$ before switching back to the initial Hermitian medium, as shown in Fig.~\ref{fig:temporalSlab}(d). Instantaneous material responses for both permittivity and conductivity have still been assumed here.

By introducing a general temporal scattering matrix formulation, which may be powerfully applied to any multi-layer configuration of temporal interfaces, such (temporally finite) TPT-symmetric structures have been proven to always reside in their symmetric phase, for which the scattering matrix supports unimodular eigenvalues. Interestingly, the non-orthogonal nature of wave interference in non-Hermitian media enables exotic energy exchanges, despite the fact that the overall temporal bilayer is in its symmetric phase. As a result, the total power flow carried by the waves after the switching events, $P_{tot} (t)=P^+ (t)+P^- (t)$ with $P^\pm (t)$ for the forward/backward propagating wave, can be significantly different from the incident one. By controlling the relative phase of two incident counter-propagating waves, we can reach drastically different total power flows for the same TPT-symmetric temporal bilayer. As an example, for $n_1=1$, $n_2=2-j0.2$ and $2\omega_1 \Delta t=3.5$, the temporal evolution of the normalized stored energy density (solid curves) and its two components $U_i=2\varepsilon\left[|E^+ (t)|^2+|E^- (t)|^2 \right]$ (dashed curves, proportional to the total power flow in the two waves $E^+ (t)$ and $E^- (t)$, and equal to the total stored energy in Hermitian media) and $U_c=2\mathrm{Re} \left[n(n^*-n) E^+ (t) E^- (t)^* \right]$ (circle symbols, stemming from the non-orthogonality of the counter-propagating waves in non-Hermitian media) are shown in Fig.~\ref{fig:temporalSlab}(e,f) for two different relative phases of the input waves, yielding widely different power flows after the switching events. Thus, dramatic power amplification and attenuation can be achieved in such non-Hermitian temporal bilayers as a function of the relative phase of the excitations, as a dual phenomenon to laser-absorber pairs in conventional PT-symmetric systems.

\subsection{Temporal switching in anisotropic media}
\label{switchingAnisotropic}
\begin{figure}[t]
    \centering
    \includegraphics{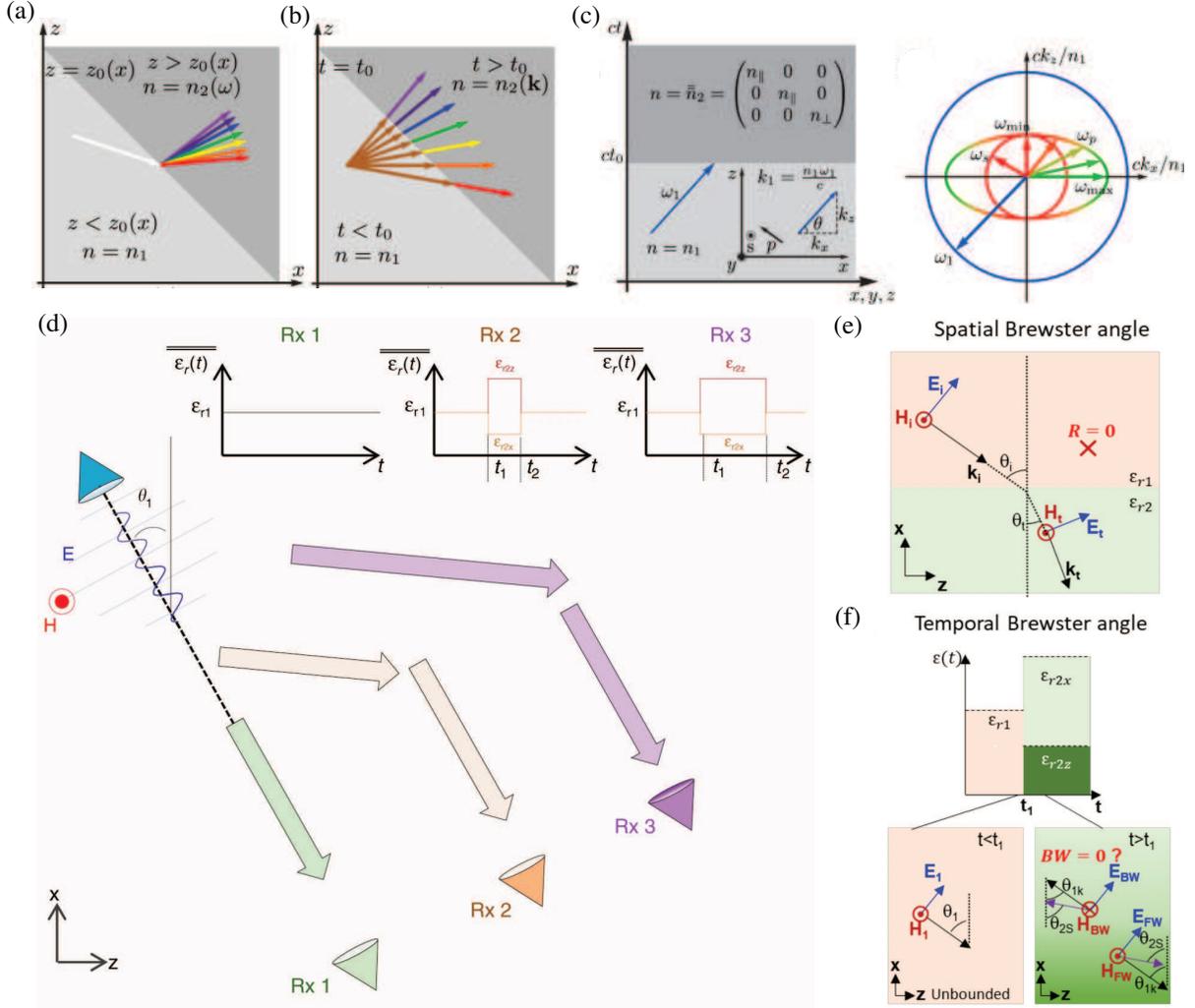}
    \caption{(a) A conventional prism decomposing white light into its frequency components in different directions. (b) An inverse prism that maps the light with different momentum into different frequencies. (c) Implementation of the inverse prism proposed in Ref.~\citeonline{akbarzadehInverse2018}. (d) Temporal aiming proposed in Ref.~\citeonline{pacheco-penaTemporal2020}. (e) The conventional Brewster angle. (f) Temporal Brewster angle described in Ref.~\citeonline{pacheco-penaTemporal2021}. Figures adapted from Refs.~\citeonline{akbarzadehInverse2018,pacheco-penaTemporal2020,pacheco-penaTemporal2021}.}
    \label{fig:switchingAnisotropy}
\end{figure}

Recent research has added new degrees of freedom to enable exotic wave transformations based on time-switching. One intriguing possibility is to consider material anisotropy and spatial dispersion. In 2018, Akbarzadeh \textit{et al.} raised the question of whether it is possible to realize an analogue of Newton’s prism, illustrated in Fig.~\ref{fig:switchingAnisotropy}(a)\cite{akbarzadehInverse2018}, in a time metamaterial. To map spatial frequencies into temporal frequencies, not only temporal invariance needs to be violated, but also spatial symmetry breaking is required to bridge different momenta with different frequency channels, as shown in Fig.~\ref{fig:switchingAnisotropy}(b). The result is an inverse prism, as presented in Fig.~\ref{fig:switchingAnisotropy}(c), in which an unbounded isotropic medium with scalar refractive index $n_1$ is switched to a uniaxial medium with $\bar{\bar{n}}_2=\mathrm{diag}\left( n_\parallel,n_\parallel,n_\perp \right)$ at $t=t_0$. The right part of panel (c) represents the isofrequency curves of the medium before and after switching. Because the medium remains homogeneous, the conservation of wavenumber $\mathbf{k}$ (momentum) still holds, leading to “birefringence” in temporal frequencies. A monochromatic wave at frequency $\omega_1$ propagating in the isotropic medium before switching will then be mapped to different frequencies, depending on its polarization and momentum, following
\begin{align}
    n_1\omega_1 = \pm n\left(\mathbf{k}\right) = 
    \begin{cases}
       \pm n_\parallel \omega_2 & s\text{-polarization  } \\
       \frac{\pm n_\parallel n_\perp \omega_2}{\sqrt{n_\parallel^2\cos^2\theta}+n_\perp^2\sin^2\theta}
       & p\text{-polarization }
\end{cases}
\end{align}
with $\theta=\tan^{-1}\left(k_z/k_x\right)$ as defined in Fig.~\ref{fig:switchingAnisotropy}(c). In general, linearly polarized light experiencing such an inverse prism will feature Lissajous polarization with a time-varying phase difference between two orthogonal field components, due to the distinct values of $\omega_2$ for \textit{s} and \textit{p}-polarizations.

One interesting phenomenon for wave propagation in anisotropic crystals is that the group velocity does not necessarily align with the phase velocity\cite{salehFundamentals2019}. Based on this feature, the concept of temporal aiming was proposed in Ref.~\citeonline{pacheco-penaTemporal2020}. Figure \ref{fig:switchingAnisotropy}(d) provides an overview of this phenomenon based on isotropic-to-anisotropic switching. A \textit{p}-polarized wave packet is immersed in a time-switched medium. Different switching schemes $\bar{\bar{\varepsilon}}_{r} (t)$ shown in the insets would route the signal to different receivers (denoted Rx 1-3  in Fig.~\ref{fig:switchingAnisotropy}(d)). For an incoming wave with propagation angle $\theta_1=\tan^{-1}\left(k_z/k_x\right)$, the direction of momentum $\mathbf{k}$ after switching to the anisotropic medium remains the same $\theta_{2k}=\theta_{1k}=\theta_1$, while that of the Poynting vector $\mathbf{S}$ is redirected to
\begin{align}
    \theta_{2S} &= \tan^{-1}\left( \frac{\varepsilon_{r2z}}{\varepsilon_{r2x}} \tan\theta_1 \right),
\end{align}

\noindent
assuming that the anisotropic medium is a uniaxial crystal with $\bar{\bar{\varepsilon}}_{2r}=\mathrm{diag}\left( \varepsilon_{2x},\varepsilon_{2x},\varepsilon_{2z} \right)$. The discrepancy between $\theta_{2S}$ and $\theta_1$ plays a key role in the aiming process by enforcing that the wave packet drifts transversely to the initial propagation direction. After an appropriate duration, the medium is then switched back to the initial state to allow the signal to travel again at the same frequency and direction. The idea of signal aiming through temporal switching paves the way to routing waves through temporal interfaces, creating a form of temporal waveguiding.

Another interesting phenomenon exploiting temporal switching with anisotropic media is the temporal Brewster angle\cite{pacheco-penaTemporal2021}. The conventional Brewster angle is defined as the angle at which the reflection of \textit{p}-polarized incidence vanishes, as shown in Fig.~\ref{fig:switchingAnisotropy}(e). Its temporal counterpart is illustrated in Fig.~\ref{fig:switchingAnisotropy}(f), and similarly determined by the condition that no backward wave is produced at the time interface. The temporal Brewster angle is given by the following simple expression:
\begin{align}
    \theta_{tB} &= \sin^{-1}\sqrt{\frac{\varepsilon_{r2x}\left( \varepsilon_{r2z}-\varepsilon_{r1} \right)}{\varepsilon_{r1}\left( \varepsilon_{r2z}-\varepsilon_{r2x} \right)}}.
\end{align}
Notice that the Brewster angle for \textit{s}-polarized waves is also expected if we consider biaxial anisotropy where $\varepsilon_{2x} \neq \varepsilon_{2y} \neq \varepsilon_{2z}$. These results open the new avenues in controlling the polarization of waves by exploiting temporal interfaces. For instance, Xu \textit{et al.} have reported complete polarization conversion using anisotropic temporal slabs\cite{xuComplete2021}.

\subsection{Temporal switching in the presence of material dispersion}

All previous results have assumed that the materials involved are non-dispersive, such that they respond to abrupt switching events instantaneously. However, caution is needed about this assumption, as the material response may have temporal dynamics comparable with the finite switching times of realistic modulation processes. In general, the temporal nonlocal response (dispersion) of a material can be considered by assuming a susceptibility kernel in the form $\mathbf{D}(t)=\varepsilon_0 \left[\varepsilon_\infty \mathbf{E}(t)+\int dt' \chi(t,t')\mathbf{E}(t-t')\right]$, where $\varepsilon_\infty$ is the background relative permittivity and $\chi(t,t')$ is the time-dependent electric susceptibility, which must satisfy causality. Recently, the generalization of the Kramers-Kronig relations have been introduced for both adiabatic\cite{hayran2021spectral} and non-adiabatic\cite{solisFunctional2021} time-varying susceptibility $\chi$. Research on temporal switching with material dispersion dates back to Fante’s and Felson’s work in 1970s\cite{fanteTransmission1971,fanteOn1973}. In parallel, wave propagation phenomena such as self-phase modulation and frequency conversion were studied in rapidly growing plasmas due to ionization\cite{yablonovitchSpectral1973,yablonovitchSelfPhase1974,jiangWave1975,wilksFrequency1988}. Comprehensive reviews on time-varying plasmas can be found in Refs.~\citeonline{mendonccaTheory2000} and~\citeonline{kalluriElectromagnetics2018}.

Once material dispersion is not negligible, it has been shown that the electric field $\mathbf{E}$ becomes continuous at a time interface, giving rise to additional temporal boundary conditions in addition to Eq. \ref{eq:time_BC}. The temporal boundary conditions in this scenario have been derived in Ref.~\citeonline{SolisTime2021} from the perspective of Parseval’s theorem and in Ref.~\citeonline{gratusTemporal2021} based on the balance of distributions. In the time domain, a Drude-Lorentz medium features a second-order differential equation for the electric polarization density $\mathbf{P}$:
\begin{align}
    \frac{d^2\mathbf{P}}{dt^2}+\gamma\frac{d\mathbf{P}}{dt}+\omega_0^2\mathbf{P} = \varepsilon_0\omega_p^2\mathbf{E},
\end{align}
where $\gamma$ is the collision damping rate, $\omega_0$ is the natural frequency of the free electron gas, and $\omega_p=\sqrt{Ne^2/\left(m_e \varepsilon_0\right)}$ with the volumetric carrier density $N$, and electron’s mass $m_e$ and charge $e$. A time-switched $N$ was considered in Refs.~\citeonline{bakunovLight2021,SolisTime2021} to effectively change $\omega_p$ abruptly. The boundary conditions at the temporal interface turn out to be the continuity of $\mathbf{B}$, $\mathbf{D}$, $\mathbf{E}$ (and therefore $\mathbf{P}$) and $d\mathbf{P}/dt$. Meanwhile, the conservation of momentum $\mathbf{k}$ in general gives rise to four solutions for the new frequencies $\pm \omega_{1,2}$.

\subsection{Time-interfaces in spatial structures}

\begin{figure}[t]
    \centering
    \includegraphics[width=\textwidth]{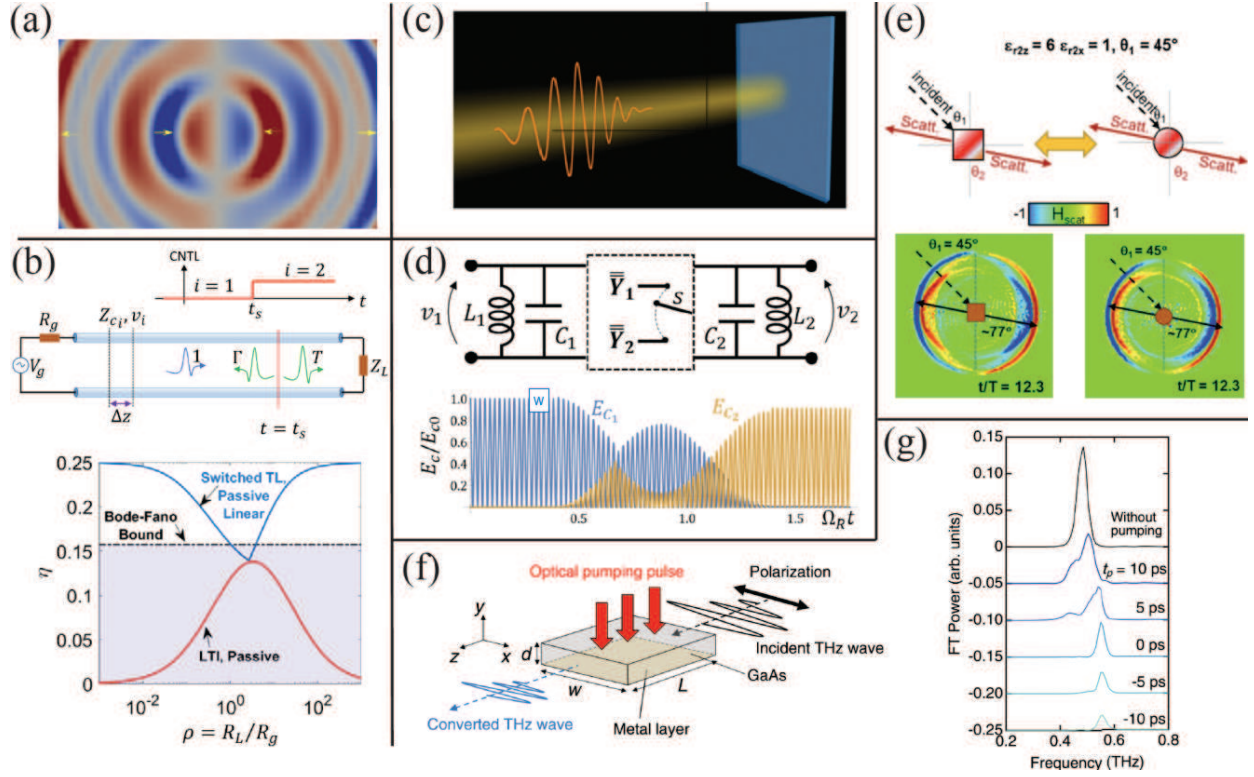}
    \caption{(a) Temporal reflection of graphene plasmons reported in Ref.~\citeonline{wilsonTemporal2018}. (b) Impedance matching using a time-switched transmission line~\cite{shlivinskiBeyond2018}. (c) Time-switched thin absorber in Ref.~\citeonline{liTemporal2021a}. (d) Unitary excitation transfer between two coupled circuit resonators~\cite{mazorUnitary2021}. (e) Temporal deflection caused by isotropic-to-anisotropic switching on meta-atoms~\cite{pachecoSpatiotemporal2021}. (f) Temporal switching of structural dispersion for ultrafast frequency-shifts, proposed in Ref.~\citeonline{miyamaruUltrafast2021}. (g) Corresponding frequency spectra for the switching in panel (f). Figures adapted from Refs.~\citeonline{wilsonTemporal2018,shlivinskiBeyond2018,liTemporal2021a,mazorUnitary2021,pachecoSpatiotemporal2021,miyamaruUltrafast2021}.}
    \label{fig:switchingFinite}
\end{figure}

While drawing significant attention, the concept of time switching still bears the question about accessibility to practical implementations. Until now experimental work demonstrating time-reversal at a temporal interface has been limited to water-wave phenomena by Fink’s group, whereby a water tank was uniformly shaken by an impulse, leading to the refocusing of the waves back to their emission point~\cite{bacotTime2016}. In addition to limitations in the achievable modulation speed and strength, one chief difficulty also lies in altering a medium in its entirety. Huge energy exchanges and stringent simultaneity of switching are typically required. Instead, more and more efforts have been geared towards switching materials only for spatially finite structures or in lower dimensions. For instance, temporal reflections were studied in Ref.~\citeonline{menendezTime2017,wilsonTemporal2018,maslovTemporal2018} for two-dimensional graphene plasmons, as shown in Fig.~\ref{fig:switchingFinite}(a). In one-dimensional transmission lines, broadband and efficient impedance matching can be achieved by time-switching, as shown in Fig.~\ref{fig:switchingFinite}(b) from Ref.~\citeonline{shlivinskiBeyond2018}. Since time-switching relaxes the constraint of time-invariance, the proposed matching schemes can go beyond the conventional Bode-Fano efficiency bounds.

Taking advantage of enhanced light-matter interactions in finite structures which support resonances, we can extend temporal switching to other applications. Switching resonant cavities enables possibilities for plasma radiation\cite{moriConversion1995} and photon-generation\cite{cironePhoton1997}. Similar ideas have also been applied to induce static-to-dynamic field radiation in a switched dielectric brick sandwiched in a waveguide\cite{mencagliStatic2021}. Li and Al{\`u} explored a time-switched Dallenbach screen, whose schematic is shown in Fig.~\ref{fig:switchingFinite}(c), showing that time-switching can extend absorption bandwidth by changing the permittivity of an absorber whilst a broadband signal enters it~\cite{liTemporal2021a}. On a related note, Mazor \textit{et al.} unveiled how an excitation can be unitarily transferred between coupled cavities by switching the coupling strength, even if the detuning between the cavities is large\cite{mazorUnitary2021}. An example of this phenomenon in a coupled LC resonator pair is shown in Fig.~\ref{fig:switchingFinite}(d): by properly switching the value of the coupling admittance, the energy stored in the capacitor $C_1$ can be transferred to $C_2$ efficiently. Subwavelength time-modulated meta-atoms have also been explored by Tretyakov's and Fleury's groups~\cite{ptitcyn2019time}, developing a consistent description of the interplay between time-modulation and dispersive polarizability~\cite{mirmoosaDipole2020}, while Engheta’s group recently extended this concept to anisotropic meta-atoms, using it to demonstrate temporal deflection\cite{pachecoSpatiotemporal2021}, as shown in Fig.~\ref{fig:switchingFinite}(e).

Finally, abrupt changes to the structural dispersion of a cavity, realized e.g. by switching part of a structure, constitute yet another intriguing degree of freedom to be exploited. For instance, in Ref.~\citeonline{miyamaruUltrafast2021}, Miyamaru \textit{et al.} achieved ultrafast terahertz frequency-shifts by metalizing the upper boundary of a single-metalized waveguide at sub-picosecond timescales via photocarrier excitation, thereby modifying its dispersion relation and inducing a large frequency shift, as shown in Fig.~\ref{fig:switchingFinite}(f,g). Similarly, nonlinear terahertz generation was reported later in ultrafast time-varying metasurfaces~\cite{tunesiTerahertz2021}. Abruptly altering a spatial boundary can be interpreted heuristically as switching the effective permittivity of the waveguide structure\cite{stefaniniTemporal2021}, which also enforces a temporal interface on the fields inside the waveguide.

To summarize, in this section we discussed wave manipulation mechanisms in the presence of temporal interfaces. We revisited the temporal boundary conditions for Maxwell's Equations and highlighted a few representative works on temporal scattering and interferences. From the theory side the field is open to the study of more sophisticated material models, such as non-Hermitian, anisotropic and even bianisotropic switching mechanisms, also in spatially structured systems. In the quest for extreme wave phenomena, temporal switching with exotic media has opened a brand-new avenue for electromagnetics research and beyond, although its ultimate impact will hinge upon further implementations of temporal switching experiments, some of which we will discuss in Sec.~\ref{sec:experiments}. 


\newpage
\section{Photonic Time-Crystals}
\label{sec:timeperiodic}

Having discussed the arsenal of potential wave phenomena based on individual time-switching, we now move on to consider systems whose properties are periodically (or quasi-periodically) modulated in time. The simplest mechanical archetype of such a parametric system is a pendulum, whose length is periodically modulated in time, one historic example being the ``Botafumeiro", an 80 kg incensory at the Cathedral of Santiago de Compostela, whose chain length is periodically modulated by monks, reaching speeds of 68 km/h over arc lengths of 65-meter within 17 modulation cycles~\cite{sanmartin1984botafumeiro} (a more familiar one is the child swing, whereby the height of the center of mass is modulated). In fact, a characteristic feature of parametric amplification is its exponential growth rate, as opposed to the linear growth that occurs in a driven oscillator.

We can make use of the scattering coefficients derived in Sec. \ref{sec:switching}, to construct the building blocks of parametric amplification in electromagnetism. After a first time-switching leading to a change of impedance $Z_1 = \sqrt{\mu_1/\varepsilon_1}\to Z_2 = \sqrt{\mu_2/\varepsilon_2}$, the displacement field at time $t$ will comprise forward and backward waves according to: $ D_2(t) \propto [T_{12} e^{j\omega_2 t} + R_{12} e^{-j\omega_1 t}] $. If we now switch the material again after a time $\Delta t_{2}$, the new forward waves will encompass doubly-reversed and doubly-transmitted waves. Conversely, the backward waves will consist of waves which were reflected on only one of the two switching events, giving, after an additional time $\Delta t_3$:
\begin{align}
    D_2/D_0 &\propto 
    (T_{12} e^{j\omega_2 \Delta t_{2}} T_{23} + R_{12} e^{-j\omega_2 \Delta t_2} R_{23}) e^{j\omega_3 \Delta t_3} \\ &+ (T_{12} e^{j\omega_2 \Delta t_2} R_{23} + R_{12} e^{-j\omega_2 \Delta t_2} T_{23}) e^{-j\omega_3 \Delta t_3} \nonumber
\end{align}
This argument can be easily extended for a given number of switching events, with forward and backward waves consisting of scattering contributions involving an even and odd number of time-reversal processes $R_{i,i+1}$ respectively.

In order to study in more detail the effect of a periodic modulation, let us set the final parameters equal to the initial ones, $\varepsilon_3 = \varepsilon_1$ and $\mu_3=\mu_1$, which forms one modulation cycle. Evaluating the energy content of the forward and backward waves, we arrive at:
\begin{align}
    |\mathbb{T}_2|^2&= 1+\frac{1}{2}\frac{(Z_1^2-Z_2^2)^2}{Z_1^2Z_2^2}\sin^2(\omega_2\Delta t_2) \\ |\mathbb{R}_2|^2&=\frac{1}{2}\frac{(Z_1^2-Z_2^2)^2}{Z_1^2Z_2^2}\sin^2(\omega_2\Delta t_2) 
    \label{eq:T2R3}
\end{align}
so that it is evident that the relation: $|\mathbb{T}_2|^2= 1+|\mathbb{R}_2|^2$ must always hold true regardless of the extent and duration of the ``temporal slab''. In order to demonstrate the generality of this argument for any number of switchings, it is instructive to keep in mind here that momentum conservation must hold true across any number of such scattering processes. Calculating either the Abraham (kinetic) momentum $\mathbf{E} \times \mathbf{H}$ or the Minkowski (canonical) momentum $\mathbf{D}\times \mathbf{B}$ must in fact yield the same total final momentum. As a result, it should not come as a surprise that any amplification of forward waves must be accompanied by an equal amplification (or generation) of backward waves, so that the overall effect of the parametric pumping is to effectively generate a standing wave on top of the incoming beam. Clearly, in a realistic scenario, since the forward and backward waves are orthogonal in the absence of modulation, they can be independently outcoupled, leading to a net increase of energy in both forward and backward waves. As another consequence of momentum conservation, the energy in the system cannot be reduced: any change in the forward wave amplitude must be compensated by the generation of a backward wave, which can only have a positive contribution to the total energy in the system. Notice how this is dual to the spatial case, where the total energy in the system must be conserved, so that the power in the forward-scattered waves can only be reduced by a scattering process. Finally, it is worth pointing out that that no energy exchange can occur from time-switching if the system is impedance-matched. As we will discuss in Sec. \ref{sec:spacetime}, however, it is still possible to have gain in impedance-matched scenarios within certain spatiotemporal modulation regimes [Sec.~\ref{sec:luminal}]. 

We now move on to study the key features of infinite time-periodic systems, which we shall regard as ``photonic time-crystals" (PTC), although it should be noted that this term bears an ambiguity with the concept of time-crystal in condensed matter physics~\cite{wilczek2012quantum}, where it denotes a stable phase of matter which spontaneously breaks continuous time-translation symmetry. Here we intend PTCs as active systems relying on an externally induced time-periodic modulation of their constitutive parameters.

In a conventional photonic space-crystal (PSC) it is well-known that Bragg scattering between waves separated in momentum by integer multiples of the reciprocal lattice vector $\mathbf{g}$ of the crystal leads to the formation of band-gaps in energy, in correspondence with the high-symmetry points in reciprocal space (in 1D, this would happen at $k=\pm g/2$). In these gaps, the states have an imaginary momentum component, thus decaying into the crystal (growing states are forbidden by energy conservation). It is expected, therefore, that modulating a system periodically in time at frequency $\Omega$ will yield band gaps in momentum ($k$-gaps) near the frequency $\Omega/2$, where $\Omega$ plays the role of a reciprocal lattice vector, as a result of the interference between the waves which are forward scattered by the modulation, and those which are time-reversed. However, while in a PSC the wavevector is imaginary inside the band-gap, in a PTC it is the energy that acquires an imaginary component within a k-gap. As expected from our discussion on momentum and energy conservation in the previous section, while waves must spatially decay into the PSC to preserve energy, they can only grow in time as they interact with a PTC, in order to preserve momentum.   
\begin{figure}[t]
    \centering
    \includegraphics[width=\textwidth]{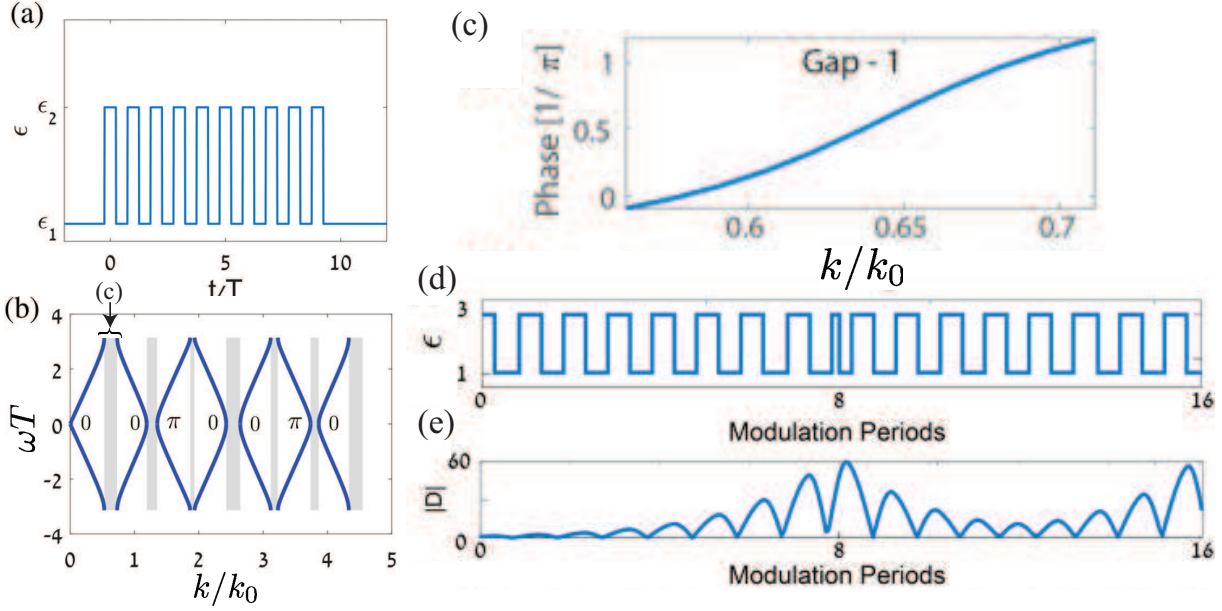}
    \caption{(a) Example of a (finite) layered PTC. (b) Band structure of a PTC, showing k-gaps and Zak numbers of the different bands. (c) Change of Zak phase across the first k-gap. (d) Example of an edge between two PTCs with opposite Zak phase. (e) Numerical demonstration of the effect of a temporal edge between two topologically inequivalent PTCs: the parametric amplification process is interrupted and the wave amplitude depleted for a few periods following the temporal edge, forming a temporally localized state. Figures reproduced from Ref.~\citeonline{lustigTopological2018}.} \label{fig:photonicTimeCrystals}
\end{figure}
Let us consider an infinite series of switching events between $\varepsilon_1$ and $\varepsilon_2$, as shown for a few cycles in Fig.~\ref{fig:photonicTimeCrystals}(a). Being the problem periodic, the Floquet (Bloch) theorem holds, so that the solution may be written as $\psi(t) = e^{i\Omega t} \phi(t)$, where $\phi(t) = \phi(t+T)$, and $T = 2\pi/\Omega$ is the period of the modulation. For the specific case of a step-like modulation, the temporal scattering coefficients can be related to each other by exploiting the symmetry $T_{12}+R_{12} = T_{21}+R_{21} = 1$, so that the Floquet bands can be calculated analytically~\cite{lustigTopological2018}, and can be seen in Fig.~\ref{fig:photonicTimeCrystals}(b). Notice how the k-gaps hosting the amplifying states open at $\omega = \Omega/2$ (i.e. $\omega T = \pi$, where $T$ is the modulation period). The reason for this can be read off the solution for the double-switching problem in Eq. \ref{eq:T2R3}. Notice how the duration $\Delta t_2$ determines whether or not gain will occur: if the duration of the modulation is half of the period $\Delta t_2 = T/2$, and we assume $\omega_1\approx\omega_2 = \omega$, then at each switching process a net energy input from the contribution of the $\sin^2(\omega T/2)$ term will be coherently fed into the waves, leading to exponential growth or, in in the present of dominant losses, loss-compensation~\cite{torrent2018loss}. Related ideas were recently investigated, such as ``multilayer" temporal slabs to design temporal transfer functions, in analogy with multilayer matching filters in spatial transmission line theory~\cite{ramaccia2021temporal,castaldi2021exploiting}, and the effects of modulating both dielectric and magnetic parameters in a PTC~\cite{gaxiola2021temporal}.

\subsection{Topology in photonic time-crystals}


One recent direction emerged in the context of spatial crystals is that of topologically nontrivial photonic phases and symmetry-protected edge modes found at their interface with a topologically inequivalent crystal. The natural question of whether such a framework exists for temporal crystals has recently been answered in the affirmative, although experimental observations of temporal topological edge states are still missing~\cite{lustigTopological2018}. In 1D (infinite) crystals the topological character of the system is quantified via the Zak phase. A temporal edge state must therefore lie at the temporal interface between PTCs characterized by different Zak phases. For a layered PTC, the Zak phase can be calculated analytically via Floquet theory, and its change across the first k-gap is shown in Fig.~\ref{fig:photonicTimeCrystals}(c)~\cite{lustigTopological2018}. Far from a trivial generalisation, however, a temporal edge is a markedly different object than a spatial one. The edge modes commonly found at the interface between topologically inequivalent crystals are now to be sought near (more specifically later than) a specific instant of time, at which the properties of the PTC are suddenly changed. In addition, while spatial edge states occur within the frequency gap of a material where the eigenstates of the infinite crystal are evanescent, temporal edge states in a PTC are only found within k-gaps, so that the underlying wave dynamics is parametric amplification, and the transient wave which constitutes the edge state has the effect of temporarily opposing the exponential growth, as shown in Fig.~\ref{fig:photonicTimeCrystals}(d-e)~\cite{lustigTopological2018}. 

\begin{figure}[t]
    \centering
    \includegraphics[width=\textwidth]{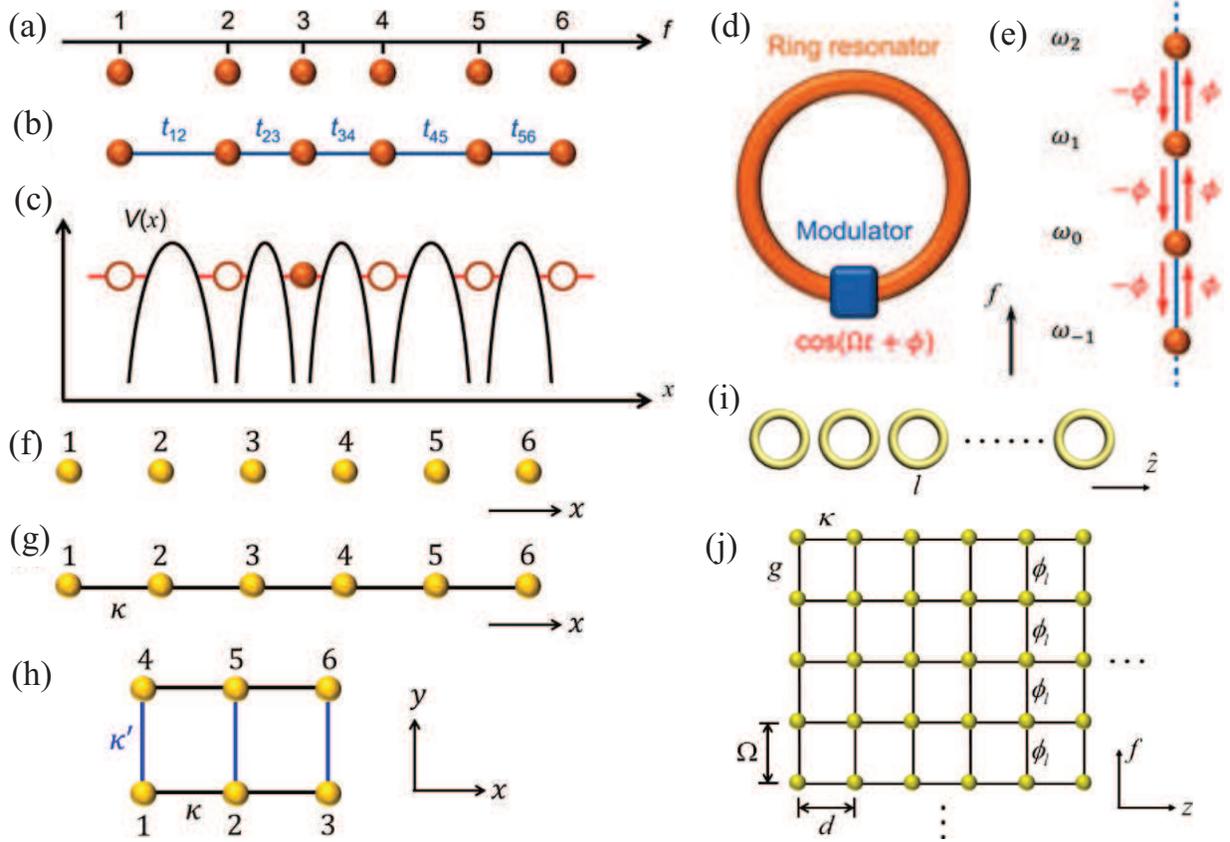}
    \caption{(a) The resonant modes of a structure form a lattice. (b) Hopping through its sites can be enabled by time modulation, which can couple their different frequencies. (c) The resulting Hamiltonian is analogous to that of electrons in a periodic ionic potential. (d) A ring resonator combined with a modulator can reproduce such a model, (e) enabling hopping between its equally spaced resonant modes~\cite{yuan2021tutorial}. (f-h) A synthetic dimension may be constructed in a tight-binding model by introducing long-term coupling between the elements of a chain. In the time-domain, this can be realized by modulating the system at multiple frequencies~\cite{yuan2018synthetic}. (i-h) A spatially 1D chain of ring resonators can be combined with temporal modulation to form a synthetic 2D lattice~\cite{yuan2016photonic}. Figures adapted from Refs.~\citeonline{yuan2016photonic,yuan2018synthetic,yuan2021tutorial}. } \label{fig:syntheticDimensions}
\end{figure}

In spite of time being a one-dimensional quantity, time-modulation can also provide a pathway towards higher-dimensional topological effects. This can be accomplished by constructing synthetic frequency dimensions, by using the frequency spectrum of the modes supported by a structure as an effective lattice of states, in analogy with the sites of a tight-binding lattice, as illustrated in Fig.~\ref{fig:syntheticDimensions}(a-c). 
Similarly to electrons in a simple one-dimensional lattice, photons in e.g. a periodically modulated ring resonator may be described by a tight-binding Hamiltonian:
\begin{align}
    H_{1} = g\sum_m (a_{m+1}^\dagger a_m e^{i\phi} + a_m^\dagger a_{m+1} e^{-i\phi})
\end{align}
where $a_m^\dagger$ and $a_m$ are the photon creation and annihilation operators for the $m^{th}$ mode on the lattice, $g = \kappa \Omega/2\pi $ is the hopping amplitude, $\kappa$ is the modulation strength, $\Omega$ is the modulation frequency and $\phi$ is the modulation phase~\cite{yuan2021tutorial}. In fact, just as electrons hop between sites of a lattice, a dynamical modulation can make photons hop between different modes of the resonator, as long as the modulation frequency is similar to the difference $\Omega_R = 2\pi v_g/L$ between the modes of the structure, where $v_g$ and $L$ are the group velocity of the waveguide constituting the resonator and $L$ is its length. The simplest instance of such a lattice may be realized with a single ring resonator combined with a periodic modulator, shown schematically in Fig.~\ref{fig:syntheticDimensions}. 

Importantly, such a ladder does not need to be unidimensional: additional dimensions may be constructed from any degree of freedom by simply introducing additional coupling terms between a lattice of resonators, or, in a synthetic frequency dimension, between modes~\cite{yuan2016photonic}, which means that a linear lattice can now be equivalently described as a folded one to highlight its higher dimensionality, as illustrated in Fig.~\ref{fig:syntheticDimensions}(f-h). With time-modulation, this can be done by considering two periodic modulations of different frequencies e.g. $\Omega_1 = \Omega_R$ and $\Omega_N = N\Omega_R$, thus introducing coupling between $N^{th}$-neighbouring modes in the lattice so that the Hamiltonian above can be generalized to:
\begin{align}
    H_{N} = g\sum_m (a_{m+1}^\dagger a_m e^{i\phi} + h.c.) + g_N\sum_m (a_{m+N}^\dagger a_m e^{i\phi_N} + h.c.)
\end{align}
where $g_N = \kappa_N\Omega_R/2/pi$. Here, the second term accounts for the coupling between modes separated by a frequency gap $N\Omega_R$, and it is mathematically equivalent to the addition of a second dimension. Time-modulation can also be combined with other degrees of freedom, such as orbital angular momentum~\cite{dutt2020single} or spatial lattices~\cite{yuan2016photonic} to implement a range of photonic effects, from non-trivial Chern numbers to effective gauge fields and spin-momentum locking. Implementations of synthetic frequency dimensions have been realized in RF with varactors~\cite{peterson2019strong} and superconducting resonators~\cite{hung2021quantum}, and in the near-infrared with fiber and on-chip electro-optic modulators~\cite{dutt2018experimental,zhang2019electronically}, as well as all-optically via four-wave mixing, with frequency spacings generally ranging from MHz to THz~\cite{yuan2021tutorial}. A recent proposal makes use of a single ring-resonator coupled to an array of optomechanical pumps and a single electro-optic modulator to realize an on-chip 2D topological insulator with two synthetic dimensions~\cite{ni2021topological}.

\subsection{Non-Hermiticity and disorder}

Another promising direction for new opportunities in wave manipulation is the interplay between temporal structure and Hermiticity. Gain in electromagnetism has conventionally been associated with the use of active media, providing wave amplification in the form of a negative dissipation. However, as discussed above, time-modulation offers an alternative route towards gain. Nevertheless, a complete analogy between non-Hermitian gain/loss and parametric processes cannot be drawn: in fact, it is evident from our previous discussion that a mere time-modulation cannot provide loss, as a consequence of momentum conservation, whereas an imaginary component in the response parameters of a medium can be used to provide either gain or dissipation depending on its sign.

\begin{figure}[t]
    \centering
    \includegraphics[width=\textwidth]{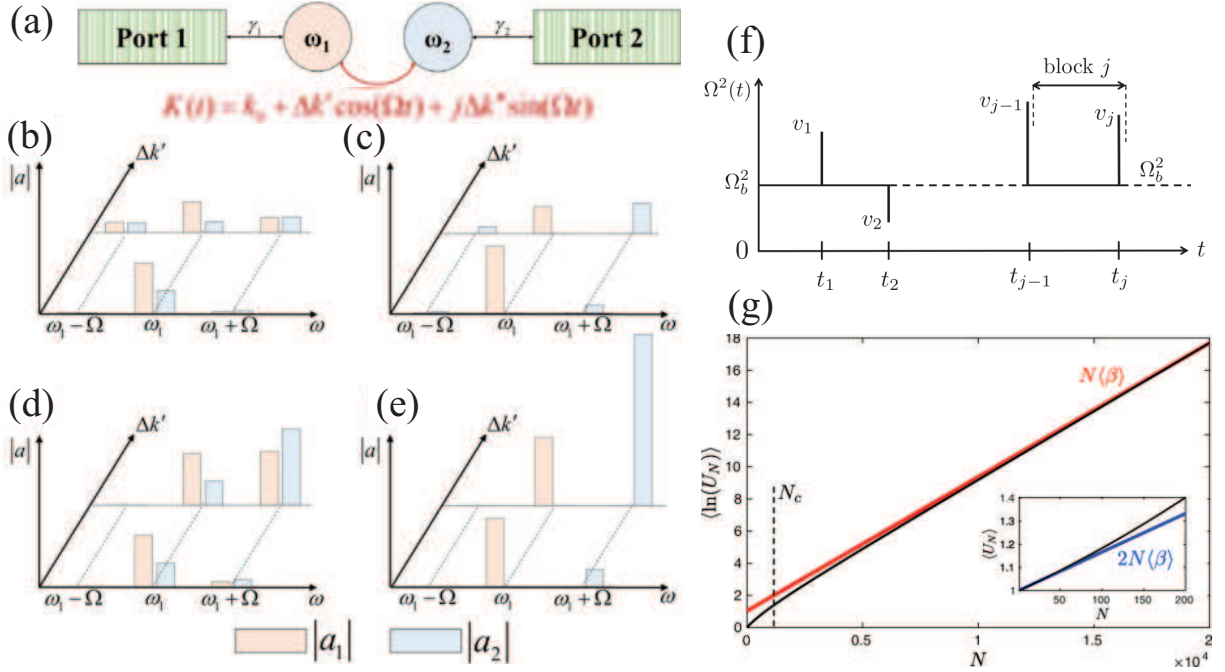}
    \caption{(a) Two resonators coupled via a combination of constant ($k_0$), time-dependent Hermitian ($\Delta k'$) and time-dependent non-Hermitian ($\Delta k''$) coupling terms. (b-e) Wave amplitude under excitation via Port 1 in the two resonators (orange and blue) for two arbitrary values of $\Delta k'$ at the excitation frequency ($\omega_1$) and the two sidebands for (b) $k_0\neq 0$ and $\Delta k''=0$ (energy stored in both sidebands and both resonators); (c) $\Delta k''=0$ and $k_0 = 0$ (energy channeled to sideband frequencies is all in the second resonator while the first only hosts the input frequency); (d) $\Delta k'=\Delta k''$ and $k_0 \neq 0$ (energy gained is nonreciprocally channeled into the upper sideband and it is distributed over both resonators); (e) $\Delta k' = \Delta k''$ and $k_0=0$ (energy gained is nonreciprocally channeled only into the upper sideband, and uniquely in the second resonator).~\cite{koutserimpas2018nonreciprocal} (f) A random sequence of temporal $\delta$-kicks results in (g) universal statistics observed by the energy in the system $U$. The energy at the N-th kick $U_N$ is plotted against the step number $N$.~\cite{carminati2021universal}. Figures adapted from Refs.~\citeonline{koutserimpas2018nonreciprocal,carminati2021universal}.} \label{fig:nonHermitianAndDisordered}
\end{figure}
This distinction between parametric and non-Hermitian gain makes it natural to ask what more can be achieved by modulating in time the non-Hermitian component of the material response. As discussed in Sec. \ref{sec:switching}, the switching of dissipation not only impacts the amplitude of the wave, but also its phase, and can also generate backward waves. In fact, as recently shown, non-Hermitian time-modulation can be used for nonreciprocal mode-steering in frequency space~\cite{koutserimpas2018nonreciprocal}. Consider the setup shown in Fig.~\ref{fig:nonHermitianAndDisordered}a: two resonators with mode-lifetimes $\gamma_1$, $\gamma_2$ are coupled via a time-dependent coupling constant $k_0+\Delta k'\cos(\Omega t) + j \Delta k'' \sin(\Omega t)$, resulting in a non-Hermitian time-Floquet Hamiltonian.  Normally a periodic modulation of the system enables similar upconversion and downconversion between modes of a structure. However, it was shown that the inclusion of a non-Hermitian component in the time-modulation of the coupling constant between the resonators can hijack such mode coupling, rerouting all of the energy into one of the two resonators, thus producing nonreciprocal gain~\cite{koutserimpas2018nonreciprocal}. 

The link between time-modulation and parity-time (PT) symmetric Hamiltonians has also been discussed in some depth for the parametric resonance case, where the frequency of the modulation doubles that of the incoming light~\cite{koutserimpas2018parametric}. Elegantly, PT symmetry breaking in these parametric systems can be related to the stability conditions of the Mathieu equation governing a harmonically modulated structure~\cite{arscott1968theory}. Such time-modulation-induced PT symmetric physics additionally enables not only lasing instabilities, but also bidirectional invisibility due to anisotropic transmission resonance, when two PT-symmetric time-Floquet slabs are combined together such that their respective parametric oscillations cancel out exactly. New perspectives relating exceptional point physics and parametric modulation may be also found in Refs.~\citeonline{wang2018photonic,bossart2021non,koutserimpas2020electromagnetic}, whereas Ref.~\citeonline{li2019topological} investigates topological phases in non-Hermitian Floquet photonic systems.

Another interesting direction for non-Hermitian time-modulated systems relates to the temporal analogue of the causality relations of conventional passive media~\cite{hayran2021spectral}. As a well-known consequence of causality, imposed by demanding that the response function of a system is only non-zero at times following an input signal, the real and imaginary parts of the response of a system must relate to each other through the Kramers-Kronig relations. However, the spectral analogue of these relations can also be constructed: if one would like the spectral response of a system to be uniquely non-zero for frequencies higher or lower than the input frequency, such that only upconversion or downconversion occurs, then there exist temporal Kramers-Kronig relations which can provide a recipe for how the Hermitian and the non-Hermitian components of a time-modulation must be related, in order for such asymmetric frequency conversion to occur. Related concepts have also been exploited for event-cloaking, broadband absorption and synthetic frequency dimensions~\cite{hayran2021spectral}.

The introduction of temporal disorder constitutes yet another avenue where the inequivalence between space and time may bear new perspectives in long-standing problems such as Anderson localization~\cite{apffel2021time,carminati2021universal,sharabi2021disordered,pendry1982evolution}. Interestingly, however, the conventional localization observed in spatially disordered media does not occur in a time-modulated system. Instead, in a similar fashion as the occurrence of edge states, the onset of the temporal analogue of Anderson localization manifests itself as an exponential growth in the energy of the waves. Such wave dynamics has been investigated both analytically~\cite{carminati2021universal}, demonstrating the universal statistics occurring in $\delta$-kicked temporal media, and numerically~\cite{sharabi2021disordered}, as a disordered perturbation in a PTC. Experiments on wave dynamics in temporally disordered systems have also recently been performed with water waves~\cite{apffel2021time}.

In the following section, we move on to briefly discuss temporally structured surfaces, whereby the interplay of spatial and temporal discontinuities can enable not only several realistic platforms for implementations with metasurfaces, but also completely new regimes of wave scattering.

\subsection{Time-modulated surfaces}

\begin{figure}[t]
    \centering
    \includegraphics[width=\textwidth]{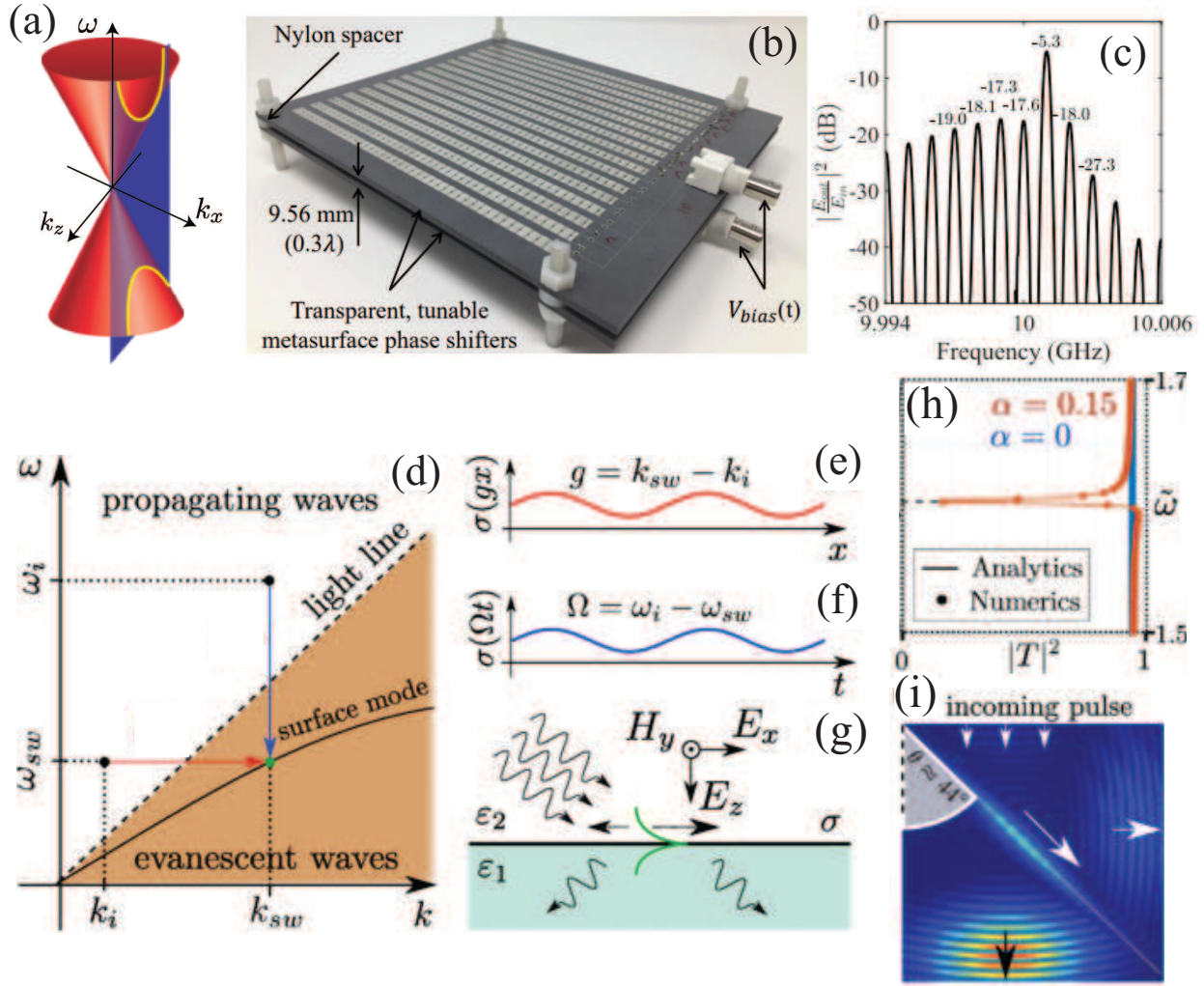}
    \caption{(a) Light scattered from a flat, time-varying metasurface can access a locus of states which form a hyperbola in phase space, as opposed to an ellipse in spatial metasurfaces. (b) Implementation of a time-modulated metasurface at GHz frequencies. (c) A 1 MHz modulation performs efficient frequency conversion through time-modulation~\cite{wu2019serrodyne}. (d) Illustration of surface-wave excitation via Wood anomalies using a spatial (e) and temporal (f) modulation of a surface (g). (h) Example of Wood anomaly observed in transmission, the surface-wave excitation is explicitly shown in (i) from a Finite-Element-Time-Domain simulation~\cite{galiffi2020wood}. Figures adapted from~\cite{wu2019serrodyne,galiffi2020wood}.}
    \label{fig:wood}
\end{figure}
Amidst the rise of 2D materials, spatially structured surfaces, known as metasurfaces, have recently attracted enormous interest as the ultrathin counterpart of metamaterials. Periodically structured surfaces, or gratings, have however occupied a central role in photonics since its early days. In its interaction with waves in the far-field, a 1D-grating can couple waves whose dispersion lives on the edge of the light cone by trading in-plane ($k_x$) and out-of-plane ($k_z$) momentum via both its in-plane structure and its transverse boundary, such that they must add in quadrature to the total momentum $k_0 = \varepsilon\omega/c_0$ of the impinging waves. Interestingly, time-modulated surfaces enable sharply different wave dynamics due the fact that the light cone treats time on a different footing that any of the spatial dimensions: if we consider light impinging on a flat, time-modulated surface, its in-plane momentum $k_x$ will now be conserved, whereas $k_z$ and, importantly the wave frequency $\omega$, are now allowed to change. Taking a vertical cut of the light cone for a fixed value of $k_x$, we see that the two quantities being traded by the surface, $k_z$ and $\omega$ must not add, but rather subtract in quadrature. As a result, the locus of points described by the available modes in the far-field is no longer elliptical, but hyperbolic. One consequence is that an increase in frequency caused by a temporal modulation corresponds to an increase in out-of-plane momentum and vice versa, the opposite of what happens between momentum components as they interact with a spatial grating. 

One direction for time-modulated surfaces is the opportunity to couple radiation to surface waves in the absence of any surface structure. Since their dispersion curve lies outside of the light cone, surface waves are typically excited by breaking translational symmetry along the surface either via a localized near-field probe like a SNOM tip~\cite{fischer1989observation}, or periodically using a grating which induces a Wood Anomaly~\cite{wood1902xlii}. However, the argument above can be exploited to envision surface-wave excitation via a temporal grating~\cite{galiffi2020wood}, as depicted in Fig.~\ref{fig:wood}(d-i). This approach brings the advantage of complete reconfigurability and use of higher-quality pristine materials, in particular in highly tunable 2D materials such as graphene and other Van der Waals polaritonic materials such as hexagonal boron nitride and MoO3, thereby circumventing the very need for surface fabrication or near-field excitation techniques.

Another key technological application of time-modulated surfaces is that of frequency conversion~\cite{cumming1957serrodyne}. The advent of metasurfaces across the electromagnetic spectrum has led to a surge of opportunities for advancing this known field of research both theoretically~\cite{wang2020theory,ramaccia2019phase,scarborough2021efficient,wu2020space,wang2021space,ramaccia2020electromagnetic,sedeh2020time,hadad2015space,romero2018parametric} and experimentally~\cite{wu2019serrodyne,zhang2021wireless,zhang2018space,shaltout2019spatiotemporal,Shaltout2019ScienceSpatiotemporal}. Recent implementations of frequency translation with metasurfaces have been recently carried out in the microwave regime~\cite{wu2019serrodyne}, and implementations with experimentally more practical discretized arrays of time-and spacetime-modulated reactive elements are in steady development~\cite{wu2020space,zhang2018space}. Several more experimental implementations are discussed in Sec.~\ref{sec:experiments}. Interestingly, the combination of bianisotropy and time-modulation has also been predicted to give rise to nonreciprocal response (see Sec.~\ref{sec:nonreciprocity})~\citeonline{wang2020nonreciprocity}. Specific scattering studies have also been dedicated to periodically modulated slabs~\cite{martinez2016temporal}, particularly in the context of parametric amplification~\cite{romero2018parametric}.

This recent surge of interest in time-modulated media has certainly highlighted the fundamental nature of the novel wave phenomena that time-modulation can unlock, and it is likely that a wealth of opportunities still lies unexplored for even the simplest spatially homogeneous systems. More opportunities involving surface structures have been explored in the context of space-time metasurfaces (Sec. \ref{sec:spacetime_metasurfaces}). However, before discussing in detail the several new related concepts and implementations, it is instructive to extend the theoretical background presented so far to the case of space-time modulations, which introduce a wealth of additional phenomenology to the scattering processes discussed so far. 

\newpage

\section{Spatiotemporal metamaterials}
\label{sec:spacetime}

\begin{figure}[t]
    \centering
    \includegraphics[width=\textwidth]{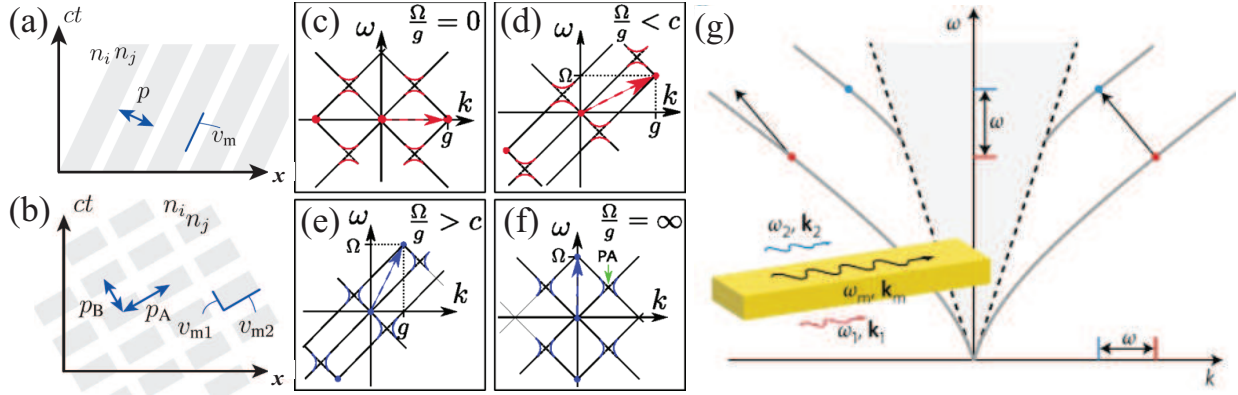}
    \caption{(a) Spacetime diagram of a space-time crystal with alternating refractive indices $n_i$ and $n_j$, spatiotemporal lattice vector $p$ and velocity $v_m$. (b) Double-period space-time crystal, characterized by two different lattice vectors $p_A$ and $p_B$, and two different velocities $v_{m1}$ and $v_{m2}$~\cite{ZoeAdvancedPhotonics}. (c) The band structure for a spatial crystal with $\varepsilon\sim \cos(gz-\Omega t)$ becomes asymmetric as (d) a finite temporal component $\Omega$, and therefore a finite modulation velocity $v_m=\Omega/g$, is introduced. This is caused by reciprocal lattice vectors (dashed arrows) acquiring a non-zero frequency component. The angle formed by the reciprocal lattice vectors with the slope of the bands determines a subluminal (c-d) and a superluminal (e-f) regime~\cite{galiffiBroadband2019}. (g) Nonreciprocal mode-coupling in a space-time modulated waveguide: the modulation can couple forward modes between each other as their frequencies and momenta are matched by the space-time modulation, while two backward waves are not coupled, so that an impinging backward wave is effectively unchanged~\cite{sounas2017non}. Figures adapted from~\citeonline{ZoeAdvancedPhotonics,galiffiBroadband2019,sounas2017non}.}
    \label{fig:basic_concepts}
\end{figure}

Spatiotemporal modulations generally imprint a travelling-wave-like perturbation, of the form $\delta\varepsilon(x,t) = \delta\varepsilon(gx-\Omega t)$ (and similarly for $\mu$) onto the response parameters of a medium, where $g=2\pi/L$ and $\Omega = 2\pi/T$ are the spatial and temporal components of the reciprocal lattice vector associated with the spatial and temporal periods $L$ and $T$ of the modulation, which form a spatiotemporal lattice vector $p=(T,L)$, as depicted in Fig.~\ref{fig:basic_concepts}(a-b)~\cite{deck-legerUniformvelocity2019}. The idea of travelling-wave modulation has been in use since the early days of travelling-wave amplifiers, whereby electron beams are used to pump energy into co-propagating microwave beams~\cite{pierce1947theory}, and have undergone a number of investigations in the 60's~\cite{cullenTravellingWave1958,olinerWave1961,Cassedy1963,cassedyDispersion1967,fante1972optical}, and later in the early 2000's~\cite{winnInterband1999,biancalanaDynamics2007}. Recently, the interest in these systems has revamped, with intense research efforts in the modelling and realization of space-time modulated nonreciprocal systems (see Sec. \ref{sec:nonreciprocity})~\cite{sounas2017non,taravatiNonreciprocal2017,sivan2016coupled,yuComplete2009,liraElectrically2012,calozSpacetime2020,calozSpacetime2020a}. In space-time media, neither frequency nor momentum are individually conserved, but rather a spatiotemporal Bloch vector $(\omega,k)$ forms a good quantum number, and the fields can generally be expressed as a superposition of a discrete set of Floquet-Bloch modes characterized by frequency-momentum vectors $(k+ng,\omega+n\Omega)$, where $n \in \mathbb{Z}$ as shown in Fig.\ref{fig:basic_concepts}(c-f), so that an eigenmode will have the form:
\begin{align}
    \psi(x,t) = e^{j(\omega t-kx)} \sum_n a_n e^{jn(\Omega t-gx)} \label{eq:spacetimeFourier}
\end{align}
and an eigenvalue problem may be set in terms of either $\omega(k)$ or $k(\omega)$. Importantly, these new reciprocal lattice vectors $(g,\Omega)$ are not generally horizontal (c), but may form an arbitrary angle in $\omega-k$ space, which may be smaller than the slope $c_0$ of the bands for the background medium (subluminal, panel d) or larger (superluminal, panel e), reducing to a pure temporal modulation in the limit $v_m=\Omega/g \to \infty$ as the entire medium is then effectively modulated instantaneously (panel f). Interestingly, a peculiar regime of modulation velocities:
\begin{align}
    \frac{c_0}{\sqrt{(1+\max(\delta\varepsilon))}\sqrt{(1+\max(\delta\mu))}}<v_m<\frac{c_0}{\sqrt{(1-\max(\delta\varepsilon))}\sqrt{(1-\max(\delta\mu))}} \label{eq:luminalRegime}
\end{align}
exists, where the Fourier expansion in Eq. \ref{eq:spacetimeFourier} does not converge, as first shown by Cassedy in Ref.~\citeonline{Cassedy1963}. We discuss this exotic \emph{luminal} regime~\cite{galiffiBroadband2019} in detail in Sec. \ref{sec:luminal}. 
In the next section we discuss the role of space-time modulation for the engineering of nonreciprocal scatterers, which first motivated the interest in this field. 

\subsection{Compact nonreciprocal devices without magnetic bias}
\label{sec:nonreciprocity}

The rise of space-time media was largely fuelled by the quest for magnet-free nonreciprocity that has dominated the metamaterials scene of the past decade. Nonreciprocity is the violation of the Lorentz reciprocity theorem, by which swapping source and receiver leaves the scattering properties of a medium unaltered~\cite{sounas2017non}. Nonreciprocal components such as isolators and circulators are crucial for duplex communications, enabling simultaneous transmission and reception at the same frequency on the same channel without parasitic reflections, as well as other applications such as laser protection from back-scattering and field enhancement. The common approach to breaking this symmetry of space is the use of gyromagnetic media, where the presence of a magnetic field breaks time-reversal symmetry. One common nonreciprocal effect encountered in basic electromagnetism is Faraday rotation, whereby the rotation of the polarization of a wave passing through a gyromagnetic medium depends on the direction from which the waves are impinging on the medium. One key drawback of gyromagnetic media, however, is their requirement of strong magnetic fields and large footprint for such effects to be appreciable, leading to bulky components, incompatible with CMOS technology. Time-modulation offers the opportunity of explicitly breaking time-reversal symmetry, thereby violating reciprocity without the use of strong magnetic fields~\cite{sounas2017non}.

\begin{figure}[t]
    \centering
    \includegraphics[width=\textwidth]{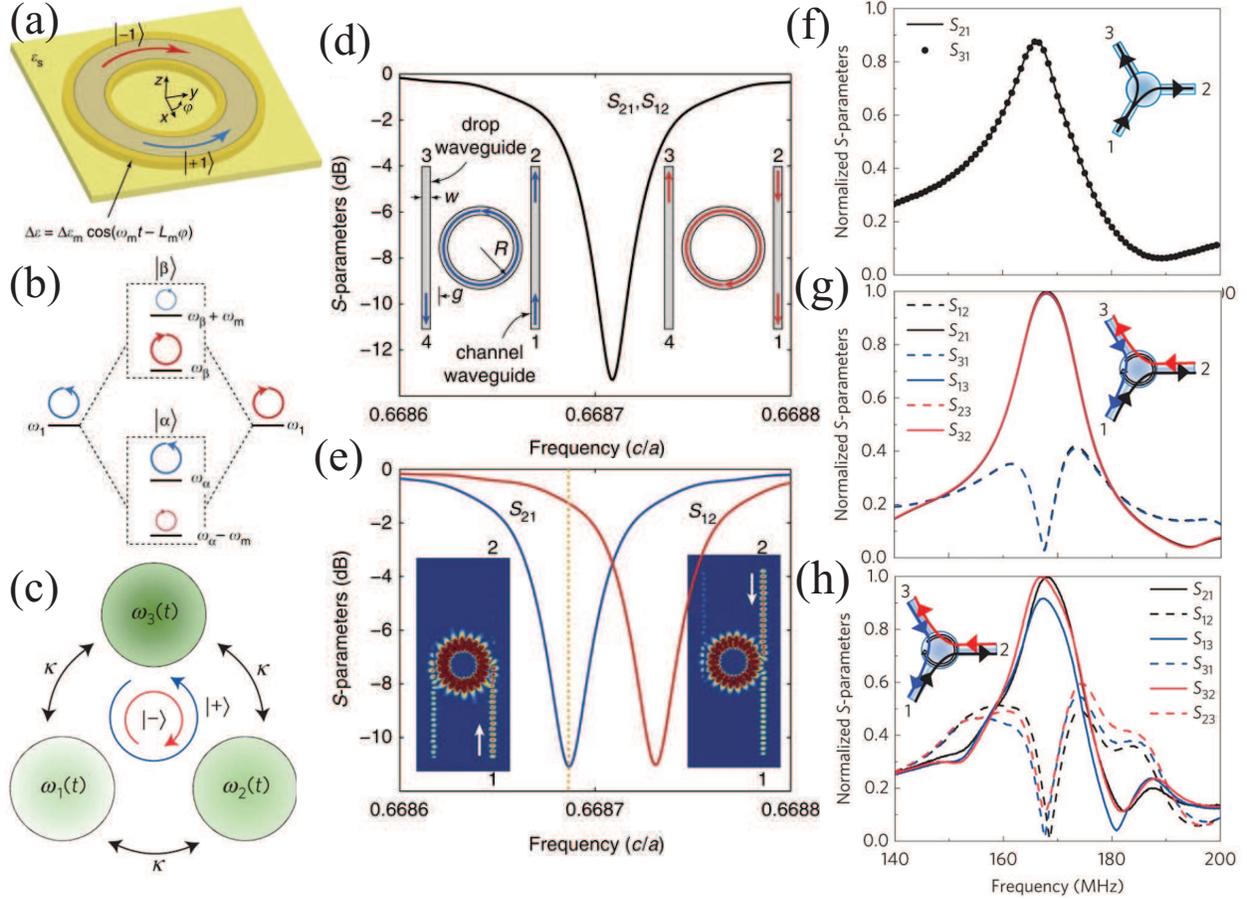}
    \caption{(a) Angular momentum biasing: a directional modulation along a ring resonator splits the degeneracy between clockwise and anti-clockwise travelling states, (b) in analogy with the Zeeman splitting of electronic states in a magnetic field~\cite{sounas2013giant}. (c) This can be realized by discretizing the elements of the ring, e.g. periodically modulating three strongly coupled resonators, with a phase of $120\deg$ between them~\cite{estep2014magnetic}. (d) Example of operation of a non-reciprocal ring resonator coupled to two waveguides: excitation from the bottom (top) of the channel waveguide (on the right of the ring) excites a counterclockwise (clockwise) mode, whose resonant frequency has been offset from the one of the clockwise (counterclockwise) mode, so that the direction of excitation determines the efficiency of the coupling to the resonator (e), and hence of the transmission to the output port of the channel waveguide~\cite{sounas2013giant}. (f-g) Theoretical and (h) experimental performance of an RF circulator made with angular-momentum biased resonators: panels (f) and (g) show the reciprocal response in the absence (f) and presence (g) of angular biasing, while panel (h) shows the performance of the experimental implementation of the circulator~\cite{estep2014magnetic}. Figures reproduced from Refs.~\citeonline{sounas2013giant,estep2014magnetic}.}
    \label{fig:nonreciprocity}
\end{figure}

The basic idea of using ST modulation for optical isolation may be summarized in Fig.~\ref{fig:basic_concepts}(g): the modulation produces transitions between two optical states which differ in both momentum $k$ and frequency $\omega$: as a result, if the system hosting the waves (e.g. a waveguide) supports multiple bands, to a pair of forward-propagating modes coupled by the ST modulation may not correspond a pair of backward-propagating ones. Therefore, while a forward incoming wave at frequency $\omega_1$ may be completely converted into a new mode with frequency $\omega_2$, a backward wave at $\omega_1$ will not be converted as it will be mismatched in frequency and/or momentum. Hence, by introducing a narrowband filter at $\omega_2$, the system allows propagation only in the backward direction, thus achieving optical isolation~\cite{sounas2017non,yuComplete2009}, in a similar fashion to the way a diode conduces electricity unidirectionally~\cite{wangOptical2013}. An equivalent way of viewing this scattering asymmetry in an extended medium is the opening of an asymmetric band gap in the photonic dispersion of a material, allowing propagation along one direction only~\cite{chamanara2017optical}.

Ring-resonators constitute perhaps the most promising components for electromagnetic nonreciprocity. The idea of using ring resonators for nonreciprocity makes use of the concept of angular momentum biasing [Fig.~\ref{fig:nonreciprocity}(a)]: a circular, directional bias splits the degeneracy between clockwise and anticlockwise states, effectively mimicking the Zeeman splitting of atomic physics [Fig.~\ref{fig:nonreciprocity}(b)]. In practice, this can be achieved with e.g. three elements periodically modulated with a $120\deg$ phase difference between them [Fig.~\ref{fig:nonreciprocity}(c)]. Let us consider the two-port system formed by two linear waveguides coupled to the biased ring resonator shown in panel d. Thanks to the splitting between clockwise and anticlockwise states, if the channel waveguide (on the right of the ring in the panel) is excited at a frequency resonant with the anticlockwise mode from the bottom port, its power will be mainly rerouted by the ring resonator into the drop channel (left of the ring). On the contrary, a wave impinging from the top of the waveguide, which would normally couple to the clockwise state would now be off-resonance, thus coupling poorly to the mode of the resonator, and being largely transmitted to the bottom port of the channel waveguide~\cite{sounas2013giant}. This strategy can be used to realize electromagnetic circulators, as shown in Fig.~\ref{fig:nonreciprocity}(f-h). A circulator consists of three ports, and its purpose is to enable transmission from port 1 to port 2, port 2 to port 3 and port 3 to port 1, while impeding transmission in the opposite sense (i.e. $2\to 1$, $3\to 2$ and $1\to 3$). Figure \ref{fig:nonreciprocity}(f) shows the transmission from channel 1 to 2 and one to 3 in the absence of angular-momentum bias, which is perfectly symmetric. Once the bias is turned on, backward propagation is forbidden, as shown in simulations [Fig.~\ref{fig:nonreciprocity}(g)] and experimental data [Fig.~\ref{fig:nonreciprocity}(h)]~\cite{estep2014magnetic}. Other electromagnetic implementations have been realized with silicon waveguides~\cite{liraElectrically2012}, as well as microstrip transmission lines~\cite{taravatiNonreciprocal2017}, with modulation frequencies ranging from hundreds of MHz to tens of GHz. 

Multiple extensive reviews on space-time modulation focusing on nonreciprocity have recently been published, so we refer the reader to them for further details, and move on to illustrate further opportunities for space-time media~\cite{sounas2017non,caloz2018electromagnetic}.

\subsection{Synthetic motion with spacetime media}

The interaction of waves with moving bodies is at the origin of a
myriad of physical phenomena, such as the Doppler effect, the Fresnel
drag by which a moving medium drags light, or even the generation of hydroelectricity~\cite{jacksonClassical2012,VanBladel,fresnel1818lettre,Fizeau}.
Moreover, the electromagnetic response of a moving system is
inherently nonreciprocal, as it is associated with a broken
time-reversal symmetry~\cite{Kong,fleurySound2014,LanneberePRA2014}. In
fact, flipping the arrow of time also requires flipping the velocity
of all the moving components, leading thereby to a distinct optical
platform. The nonreciprocity and the non-Hermitian nature of the electromagnetic response of moving matter can enable unidirectional light flows~\cite{LanneberePRA2014}, classical and quantum non-contact friction~\cite{PendryFriction,SilvFriction}, parametric amplification~\cite{Zeldovich,SilvPRX2014, LanneberePRA2014, FaccioPRL2017}, amongst others~\cite{philbin2008fiber}.

Unfortunately, it is impractical to take technological advantage of
many of the features highlighted in the previous paragraph, because
a realistic value for the velocity of a moving body is many orders
of magnitude smaller than the speed of light. Interestingly, the
actual physical motion of a macroscopic body may be imitated by a
drift current in a solid state material with high-mobility
\cite{Sydoruk2010, Morgado2017, VanDuppen2016, MorgadoACS2018}. For example, it was recently experimentally verified that drifting electrons in graphene can lead to nonreciprocity at terahertz frequencies and to a Fresnel drag~\cite{BasovNature2021,WangNature2021}. A different way to realize an effective moving response without any actual physical motion is through a travelling wave spacetime modulation~\cite{olinerWave1961,Cassedy1963,cassedyDispersion1967,deck-legerUniformvelocity2019,huidobroFresnel2019,mazor2019one}. Such a solution is discussed next in detail.

\begin{figure}[t]
    \centering
    \includegraphics[width=\textwidth]{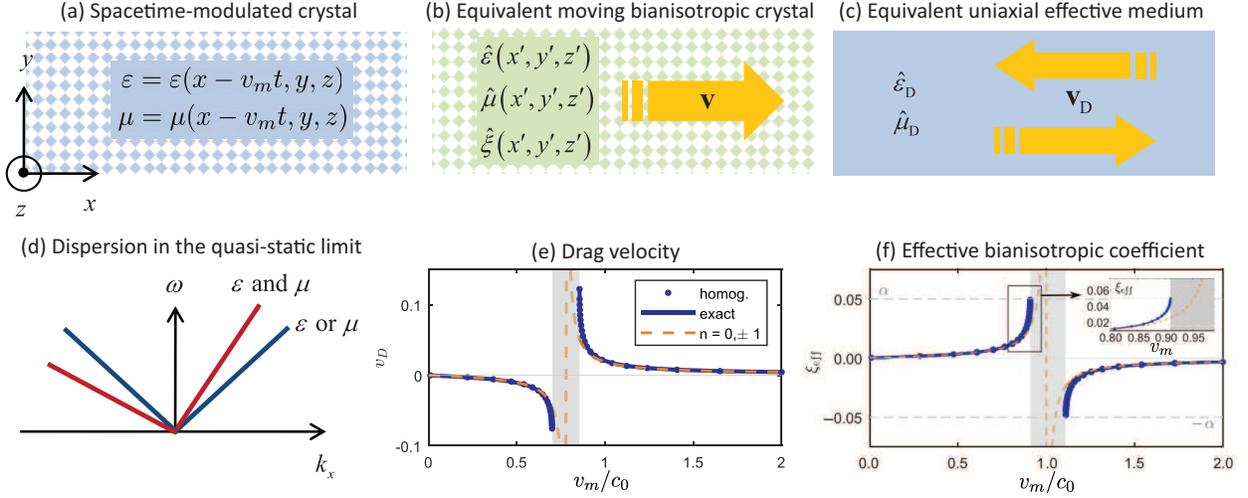}
    \caption{(a) and (b) A synthetic spacetime modulated crystal (panel a) is formally equivalent to a fictitious moving bianisotropic time-invariant crystal (panel b) when $v_m<c_0$. The velocity of the equivalent moving crystal is identical to the modulation speed $\bf{v_m}$. (c) In the long wavelength limit the system response can be homogenized
  and the crystal behaves as a uniaxial dielectric moving with speed $\bf{v_D}$. The sign of $v_D$ is not necessarily coincident with the sign of $v$.  (d) Dispersion diagram of the effective medium $\omega \,\, {\rm{vs}} \,\, k_x$. When only one of the material parameters is spacetime modulated the dispersion is symmetric, and the equivalent drag velocity vanishes (blue lines). When both $\varepsilon$ and $\mu$ are spacetime modulated, the dispersion is asymmetric (red lines, assuming $v_D>0$). The waves that propagate in the direction of $\bf{v}_D$ have a larger group velocity than the waves that propagate in the opposite direction~\cite{huidobroFresnel2019}.
  (e) Drag velocity $v_D$ and (f) effective bianisotropic coefficient $\xi_{\rm{ef}}$ as a function of the modulation speed $v_m$ for a representative spacetime modulated crystal: the velocity of the equivalent moving medium and the magneto-electric coupling coefficient flip sign in the transition from the subluminal to the superluminal regime.
  Adapted from~\citeonline{huidobro2021homogenization}.}
    \label{fig:synthetic_motion}
\end{figure}

A system featuring a travelling-wave-type modulation is characterized, in the dispersionless limit, by
the following constitutive relations:
\begin{equation}
\left( {\begin{array}{*{20}{c}}
{{\bf{D}}}\\
{{\bf{B}}}
\end{array}} \right) = \left( {\begin{array}{*{20}{c}}
{\varepsilon \left( {x - {v_m}t,y,z} \right){\bf{1}}}&{\bf{0}}\\
{\bf{0}}&{\mu \left( {x - {v_m}t,y,z} \right){\bf{1}}}
\end{array}} \right) \cdot \left( {\begin{array}{*{20}{c}}
{{\bf{E}}}\\
{{\bf{H}}}
\end{array}} \right) \label{eq:constitutiveLab}
\end{equation}
where $v$ is the modulation speed [Fig.~\ref{fig:synthetic_motion}(a)]. 
This form of modulation in space and time imparts a synthetic motion to the material response along the $x$-direction with uniform speed $v_m$.
This property implies that with a suitable coordinate transformation one can switch to a frame where the material response is time-invariant. In fact, a Galilean transformation of coordinates,
\begin{equation}
x' = x - {v_m}t, \quad y'=y, \quad z'=z, \quad t' = t,
\label{eq:GalileanT}
\end{equation}
preserves the usual structure of the Maxwell's equations, with the
transformed fields related to the original fields as follows
${\bf{D'}} = {\bf{D}}$, ${\bf{B'}} = {\bf{B}}$, ${\bf{E'}} =
{\bf{E}} + {\bf{v_m}} \times {\bf{B}}$, ${\bf{H'}} = {\bf{H}} -
{\bf{v_m}} \times {\bf{D}}$ and ${\bf{v_m}} = {v_m}{\bf{\hat x}}$. In the
new coordinates, the constitutive relations are time-independent
\begin{equation}
\left( {\begin{array}{*{20}{c}}
{{\bf{D'}}}\\
{{\bf{B'}}}
\end{array}} \right) = \left( {\begin{array}{*{20}{c}}
{{\bf{\hat \varepsilon '}}\left( {{\bf{r'}}} \right)}&{{\bf{\hat \xi '}}\left( {{\bf{r'}}} \right)}\\
{{\bf{\hat \zeta '}}\left( {{\bf{r'}}} \right)}&{{\bf{\hat \mu
'}}\left( {{\bf{r'}}} \right)}
\end{array}} \right) \cdot \left( {\begin{array}{*{20}{c}}
{{\bf{E'}}}\\
{{\bf{H'}}}
\end{array}} \right), \label{eq:constitutiveComoving}
\end{equation}
with ${\bf{r'}} = {\bf{r}} - {\bf{v_m}}t$ and the transformed
effective parameters given by
\begin{align}
    {\bf{\hat \varepsilon '}} &= \varepsilon \left( {\frac{1}{{1 -
\varepsilon \mu {v_m^2}}}{{\bf{1}}_t} + {\bf{\hat x}} \otimes
{\bf{\hat x}}} \right), \\
{\bf{\hat \mu '}} &= \mu \left( {\frac{1}{{1 - \varepsilon \mu
{v_m^2}}}{{\bf{1}}_t} + {\bf{\hat x}} \otimes {\bf{\hat x}}} \right),\\{\bf{\hat \xi '}} &=  - {\bf{\hat \zeta '}} = \frac{{ - \varepsilon
\mu }}{{1 - \varepsilon \mu {v_m^2}}}{\bf{v_m}} \times {\bf{1}}.
\end{align}
In the above, ${{\bf{1}}_t} = {\bf{\hat y}} \otimes {\bf{\hat y}} +
{\bf{\hat z}} \otimes {\bf{\hat z}}$ and $\otimes$ represents the
tensor product of two vectors. As seen, the coordinate transformation originates a bianisotropic-type coupling determined by the magneto-electric tensors $\bf{\hat \xi '}$ and $\bf{\hat \zeta '}$~\cite{Serdyukov}, such that the electric displacement vector $\bf{D'}$ and the magnetic induction $\bf{B'}$ depend on both the electric and magnetic fields $\bf{E'}$ and $\bf{H'}$. Indeed, as
further discussed below, one of the peculiar features of the travelling-wave spacetime modulation is that it mixes the electric and magnetic responses, leading to the possibility of a giant bianisotropy in the quasi-static limit~\cite{huidobro2021homogenization,huidobroFresnel2019}.

The new coordinate system is not associated with an inertial frame. An immediate consequence of this property is that the vacuum response does not stay invariant under the coordinate transformation (\ref{eq:GalileanT}). Usually, this does not create any difficulties, but it is relevant to mention that it is also possible to obtain a time-invariant response with a Lorentz coordinate transformation. Such a solution is restricted to subluminal modulation velocities, and thereby the Galilean transformation is typically the preferred one.

Interestingly, the constitutive relations (\ref{eq:constitutiveComoving}) in the new coordinate system are reminiscent of those of a moving dielectric medium~\cite{Kong}. Due to this feature the electrodynamics of spacetime media with travelling wave modulations bears many similarities to the electrodynamics of moving bodies~\cite{huidobroFresnel2019}. Yet, it is important to underline that a spacetime modulated dielectric crystal is not equivalent to a \emph{moving} dielectric crystal. In other words, impressing a time modulation on the parameters of a dielectric photonic crystal is not equivalent to setting the same photonic crystal into motion. A moving dielectric crystal would have a bianisotropic response in a frame where its speed $v$ is nontrival, very different from the constitutive relations (\ref{eq:constitutiveLab}). Below, we revisit this discussion in the context of effective medium theory.

Even though the spacetime modulated dielectric system is not
equivalent to a moving dielectric, its response can be precisely linked
to that of a fictitious moving medium in the subluminal case. In
fact, as previously noted, a suitable Lorentz transformation makes
the response (\ref{eq:constitutiveLab}) of a spacetime system
independent of time, analogous to Eq.
(\ref{eq:constitutiveComoving}). Thus, the original constitutive
relations (\ref{eq:constitutiveLab}) are indistinguishable from those
of a hypothetical moving system with a bianisotropic response in
the co-moving frame of the type (\ref{eq:constitutiveComoving}). In
other words, the synthetic motion provided by the travelling wave
modulation imitates the actual physical motion of a fictitious
time-invariant bianisotropic crystal [Fig.~\ref{fig:synthetic_motion}(b)].

The previous discussion is completely general, apart from the assumption of a travelling wave modulation. In the following, we focus our attention on periodic systems and in the long wavelength regime. Effective medium methods have long been used to provide a simplified description of the wave propagation in complex media and metamaterials~\cite{DRSmith2007, Silv2007, Simovski2007,AluFirstPrinciples2007}. The effective medium formalism is useful not only because it enables analytical modeling of the relevant phenomena, but also because it clearly pinpoints the key features of the system that originate the peculiar physics.

The standard homogenization approach is largely rooted on the idea of ``spatial averaging'', so that the effective response describes the dynamics of the envelopes of the electromagnetic fields~\cite{jacksonClassical2012}. In particular, in the traditional framework, there is no time averaging. Due to this reason, standard homogenization approaches are not directly applicable to spacetime crystals, where the microscopic response of the system depends simultaneously on space and on time.

At present, there is no general effective medium theory for spacetime crystals. Fortunately, the particular class of spacetime crystals with travelling wave modulations can be homogenized using standard ideas~\cite{huidobro2021homogenization, HTToHomogenization,LurieBook}. In fact, since a travelling wave modulated spacetime crystal is virtually equivalent to a moving system, it is possible to find the effective response with well established methods by working in the co-moving frame (primed coordinates) where the medium response is time invariant~\cite{huidobro2021homogenization}.

To illustrate these ideas, we consider a one-dimensional photonic
crystal such that the permittivity $\varepsilon$ and permeability
$\mu$ are independent of the $y$ and $z$ coordinates. It is well
known, that for stratified systems the components of the $\bf{E}$
and $\bf{H}$ fields parallel to the interfaces ($y$ and $z$
components) and the components of the $\bf{D}$ and $\bf{B}$ fields
normal to the interfaces ($x$ components) are constant in the long
wavelength limit~\cite{Aspnes1982}, i.e., when the primed field
envelopes vary sufficiently slowly in space and in time. Taking this
result into account, one can relate the spatially averaged
$\left\langle {{\bf{D'}}} \right\rangle$ and $\left\langle
{{\bf{B'}}} \right\rangle$ with the spatially averaged $\left\langle
{{\bf{E'}}} \right\rangle$ and $\left\langle {{\bf{H'}}}
\right\rangle$~\cite{huidobro2021homogenization}. The effective parameters
in the original (laboratory) frame can then be found with an inverse
Galilean (or Lorentz) transformation. Such procedure leads to the
following constitutive relations in the lab frame:
\begin{equation}
\left( {\begin{array}{*{20}{c}}
{\left\langle {\bf{D}} \right\rangle }\\
{\left\langle {\bf{B}} \right\rangle }
\end{array}} \right) = \left( {\begin{array}{*{20}{c}}
{{{{\bf{\hat \varepsilon }}}_{{\rm{ef}}}}}&{{\xi _{{\rm{ef}}}}{\bf{\hat x}} \times {\bf{1}}}\\
{ - {\xi _{{\rm{ef}}}}{\bf{\hat x}} \times {\bf{1}}}&{{{{\bf{\hat
\mu }}}_{{\rm{ef}}}}}
\end{array}} \right) \cdot \left( {\begin{array}{*{20}{c}}
{\left\langle {\bf{E}} \right\rangle }\\
{\left\langle {\bf{H}} \right\rangle }
\end{array}} \right), \label{eq:effectiveconstitutive}
\end{equation}
where ${\bf{\hat \varepsilon }}_{\rm{ef}}$,  ${\bf{\hat \mu
}}_{\rm{ef}}$, and $\xi_{\rm{ef}}$ are some parameters that can be written explicitly in terms of the permittivity and permeability profiles $\varepsilon(x)$ and $\mu(x)$~\cite{huidobro2021homogenization}. It turns out that when only one of the material parameters is
modulated in spacetime the magnetoelectric coupling coefficient
$\xi_{\rm{ef}}$ vanishes and the effective medium behaves as a
standard uniaxial dielectric~\cite{huidobro2021homogenization,huidobroFresnel2019}. For example, if the material permeability is
independent of space and time, then $\xi_{\rm{ef}}=0$ and ${\bf{\hat
\mu }}_{\rm{ef}} = \mu {\bf{1}}$, for any permittivity profile
$\varepsilon(x)$. This property can be intuitively understood by noting that when $\mu =const. $, the ``microscopic'' constitutive relation ${\bf{B}} = \mu {\bf{H}}$ implies that the averaged fields are also linked by $\left\langle {\bf{B}} \right\rangle = \mu \left\langle {\bf{H}} \right\rangle$, which corresponds to a trivial effective magnetic response. Note that this argument is only justified in the static limit, as in the dynamical case the second order spatial dispersion effects may lead to artificial magnetism~\cite{Pendry1999, Silv2007}.

Remarkably, when both $\varepsilon$ and $\mu$ are modulated in
spacetime, $\xi_{\rm{ef}}$ can be nontrivial. In other words,
notwithstanding that at the microscopic level there is no
magnetoelectric coupling [Eq. (\ref{eq:constitutiveLab})], the
effective medium is characterized by a bianisotropic response in the
static limit. While it is not unusual that the complex wave
interactions in a metamaterial without inversion symmetry can result
in a magnetoelectric coupling~\cite{Serdyukov}, having a nontrivial
bianisotropic response in the long wavelength limit is a rather
unique result~\footnote{For completeness, we point out that some
antiferromagnets, such as chromium oxide, may be characterized by an
intrinsic axion-type bianisotropic response in the static limit
\cite{Astrov, CohVanderbilt}.}. In fact, in typical time-invariant systems the electric and magnetic fields are decoupled in the static limit. Thereby, the electric and magnetic responses of any conventional metamaterial are necessarily decoupled in the static limit and there is no bianisotropy. In contrast, for a spacetime modulated system the electric and the magnetic fields are never fully decoupled, as the time modulation of the material implies always some nontrivial time dynamics. The strength and sign of $\xi_{\rm{ef}}$ can be tuned by changing the modulation speed and the profiles of the permittivity and permeability. As illustrated in Fig.~\ref{fig:synthetic_motion}(f), $\xi_{\rm{ef}}$ can be exceptionally large for modulation speeds approaching the speed of light in the relevant materials. This results in giant nonreciprocity, as has also been shown through Floquet-Bloch expansions~\cite{taravatiGiant2018}. The transition from the subluminal to the superluminal regime is marked by a change in sign of $\xi_{\rm{ef}}$. 
Furthermore, it can be shown that constitutive relations
(\ref{eq:effectiveconstitutive}) are identical to those of a
fictitious moving anisotropic dielectric characterized in the
respective co-moving frame by the permittivity ${{{\bf{\hat\varepsilon }}}_{{\rm{D}}}}$ and by the permeability ${{{\bf{\hat\mu }}}_{{\rm{D}}}}$. The equivalent medium moves with a speed $v_D$ with respect to the lab frame. The parameters ${{{\bf{\hat\varepsilon }}}_{{\rm{D}}}}$, ${{{\bf{\hat \mu }}}_{{\rm{D}}}}$, and $v_D$ are determined univocally by ${\bf{\hat \varepsilon}}_{\rm{ef}}$, ${\bf{\hat \mu }}_{\rm{ef}}$, and $\xi_{\rm{ef}}$\cite{huidobro2021homogenization}. In general, $v_D$ may be rather different from the modulation speed $v$, both in amplitude and sign. The velocity $v_D$ can be nonzero only if $\varepsilon$ and $\mu$ are both modulated in spacetime, i.e., only if $\xi_{\rm{ef}} \neq 0$. Therefore, it follows that a spacetime modulated photonic crystal can imitate precisely the response of a moving dielectric in the long wavelength limit. Remarkably, the velocity of the equivalent moving medium can be a very significant fraction of $c_0$ [Fig.~\ref{fig:synthetic_motion}(e)]. Thus, spacetime modulated systems are ideal platforms to mimic on a table-top experiment the electrodynamics of bodies moving at relativistic velocities, the only caveat being that some degree of frequency mixing will, in general, occur upon propagation in a space-time medium.

In particular, similar to a moving medium, a spacetime modulated crystal can produce a drag effect~\cite{huidobroFresnel2019}. Specifically, a wave co-propagating  with the equivalent moving medium, i.e. towards the direction ${\rm{sgn}}\left( {{v_{\rm{D}}}} \right){\bf{\hat x}}$, moves faster than a wave propagating in the opposite direction, due to the synthetic Fresnel-drag effect~\cite{Fizeau}. A clear fingerprint of the Fresnel drag can be detected in the dispersion of $\omega$ vs. $k$ of the electromagnetic modes of the spacetime crystal. In fact, when both $\varepsilon$ and $\mu$ are modulated in time, the slopes of the dispersion of the photonic states near $k = 0$ depends on the direction of propagation of the wave. The dispersion of the waves that co-propagate with the equivalent moving medium exhibits a larger slope than the dispersion of the modes that propagate in the opposite direction, as depicted in Fig.~\ref{fig:synthetic_motion}(d). Interestingly, these slopes are predicted \emph{exactly} by the discussed effective medium theory, and thereby the homogenization theory is expected to be rather accurate and useful to characterize excitations that vary sufficiently slowly in space and in time. Furthermore, the effective medium theory is exact for all frequencies when the parameters of the crystal are matched: $\varepsilon /\mu = const.$. In fact, the dispersion of a matched photonic crystal is generally linear for all frequencies. However, as we show in the next section, even this statement can be violated in an exotic class of spacetime media recently termed \emph{luminal} metamaterials, which we discuss in the next section.

\subsection{Luminal amplification and spatiotemporal localization}

\label{sec:luminal}

\begin{figure}[t]
    \centering
    \includegraphics[width=\textwidth]{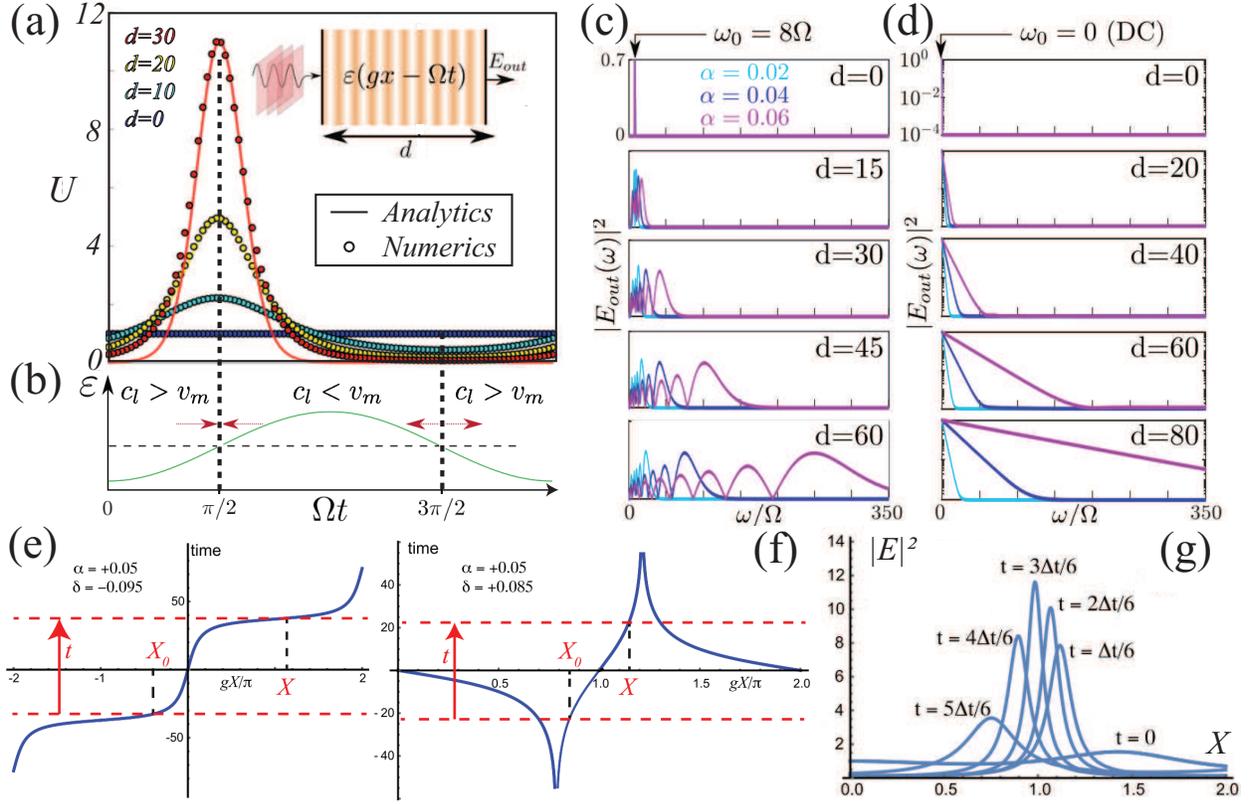}
    \caption{(a) Electric field intensity at the output port vs time transmitted through different thicknesses (blue to red) $d$ of luminal crystal. (b) In the gain region the permittivity gradient is positive, and vice versa. The gain is maximum at the points (in time or space) where the velocity of the waves matches the velocity of the grating. (c-d) Power spectra of the transmitted waves for input frequency (c) $\omega_0 = 8\Omega$, with $\Omega$ the modulation frequency, and (d) for a DC input field ($\omega_0 = 0$), showing the frequency generation responsible for the overall gain~\cite{galiffiBroadband2019,galiffi2020new}. (e-f) Trajectories of a point of constant phase along the spacetime variable $X=x-v_m t$. (e) outside of the localization regime (resulting in oscillations which periodically compress and decompress the pulse, panel g), and (f) within the localization regime, resulting in the amplification and compression described in (a-d)~\cite{galiffiPhoton2021}. Figures adapted from Refs.~\citeonline{galiffi2020new,galiffiBroadband2019,galiffiPhoton2021}.}
    \label{fig:luminal}
\end{figure}

If the speed of a space-time modulation falls within the modulation velocity regime in Eq.~\ref{eq:luminalRegime} (the extent of this \emph{luminal} regime increases with the modulation amplitude), the relevant wave physics changes dramatically, entering a very peculiar unstable phase whose underlying mechanism is, however, completely distinct from the parametric amplification processes described in Secs. \ref{sec:switching} and \ref{sec:timeperiodic}~\cite{galiffiBroadband2019}. In this regime, all forward bands become almost degenerate, as the frequency-momentum reciprocal lattice vectors effectively align with the forward bands. This regime marks a transitions between the subluminal and superluminal regimes in Fig.~\ref{fig:basic_concepts}(d) and (e) respectively. Due to the resulting strong degeneracy between forward-travelling waves, the eigenmodes of these luminal systems cannot be written in Bloch form~\cite{galiffiPhoton2021}: the effect of the modulation, in fact, is to couple all of the forward bands, such that an impinging wave would emerge as a supercontinuum, or frequency comb. Such a transmission process is shown in Fig.~\ref{fig:luminal}(a-d): the waves are compressed into a train of pulses, and both the total energy of the system and its compression increase exponentially with respect to the propagation length (or time) in the luminal medium, as well as the amplitude and rate of the modulation. In real space, this amplification process can be viewed as the capturing of field lines by the synthetically moving grating~\cite{pendry2021gain}. The wave dynamics in a luminal medium is captured by the approximate equation for the energy density $U$ (which becomes exact if $\varepsilon=\mu$ everywhere):
\begin{align}
    \frac{\partial \ln{U}}{\partial t'} = \frac{(v_m+c_l)}{2} \bigg[ \frac{\partial \ln\mu}{\partial X} + \frac{\partial \ln\varepsilon}{\partial X}  \bigg] + (v_m-c_l) \frac{\partial \ln{U}}{\partial X} \label{eq:luminalEq}
\end{align}
where $v_m$ is the grating velocity, $c_l(X) = (\varepsilon\mu)^{-1/2}$ and $X = x-v_m t$ and $t' = t$ are moving coordinates, which can be solved by successive iterations~\cite{galiffi2020new}. The characteristic feature of the luminal regime is the presence of points along the grating where the local phase velocity $c_l(X)$ of the waves matches the velocity $v_m$ of the grating (crossings between the black, dashed horizontal line and the green sinusoidal grating profile in Fig.~\ref{fig:luminal}(b)). The presence of these velocity-matching points implies that the lines of force are trapped within each period of the grating, merely subject to the first term, which acts as an energy source term, resulting from the spatiotemporal change in $\varepsilon$ and $\mu$. In fact, these local gradients in the electromagnetic parameters (first term in Eq. \ref{eq:luminalEq}) will pump energy into the waves, or deplete them, based on their sign, whereas the phase velocity of the waves in the region adjacent to the velocity-matching points determines whether these will function as attractors ($\pi/2$ in Fig.~\ref{fig:luminal}(b)) or repellors ($3\pi/2$ in Fig.~\ref{fig:luminal}(b)) as a result of the Poynting flux driving power towards or away from them\cite{pendryNew2021}. It has been noted that this completely linear mechanism results in pulse compression or expansion~\cite{chamanaraLinear2019}. On the other hand, as the modulation velocity approaches the edges of the luminal regime, a localization transition takes place, before which the response of the system is characterized by strong oscillations in time consisting of periodic field compression and expansion~\cite{galiffiPhoton2021}, as shown in figure \ref{fig:luminal} (e-g). Importantly, luminal amplification does not rely on back-scattering, as opposed to parametric amplification, and in fact it occurs just as well in a system which is impedance-matched everywhere in spacetime, in sharp contrast to parametric amplification, which relies on time-reversal as explained in Sec. \ref{sec:timeperiodic}. 


\newpage
\subsection{Space-time metasurfaces}
\label{sec:spacetime_metasurfaces}
Most implementations of time-varying and space-time-varying systems are hard to achieve in bulk, as any modulation mechanism, e.g. a pump pulse, would generally need to impinge on the system from an additional direction. In addition, field penetration into a bulky structure is generally inhomogeneous, so that in practice only a small surface layer of a bulk medium would effectively be modulated. These considerations, combined with with the ease of fabrication of metasurfaces compared to bulk metamaterials, have led to a concentration of research interest in space-time metasurfaces, whereby a travelling-wave modulation is applied to the surface impedance of a thin sheet, with proposals covering a wide range of ideas and applications, some of which are shown in Fig.~\ref{fig:nonreciprocity}.

\begin{figure}[t]
    \centering
    \includegraphics[width=\textwidth]{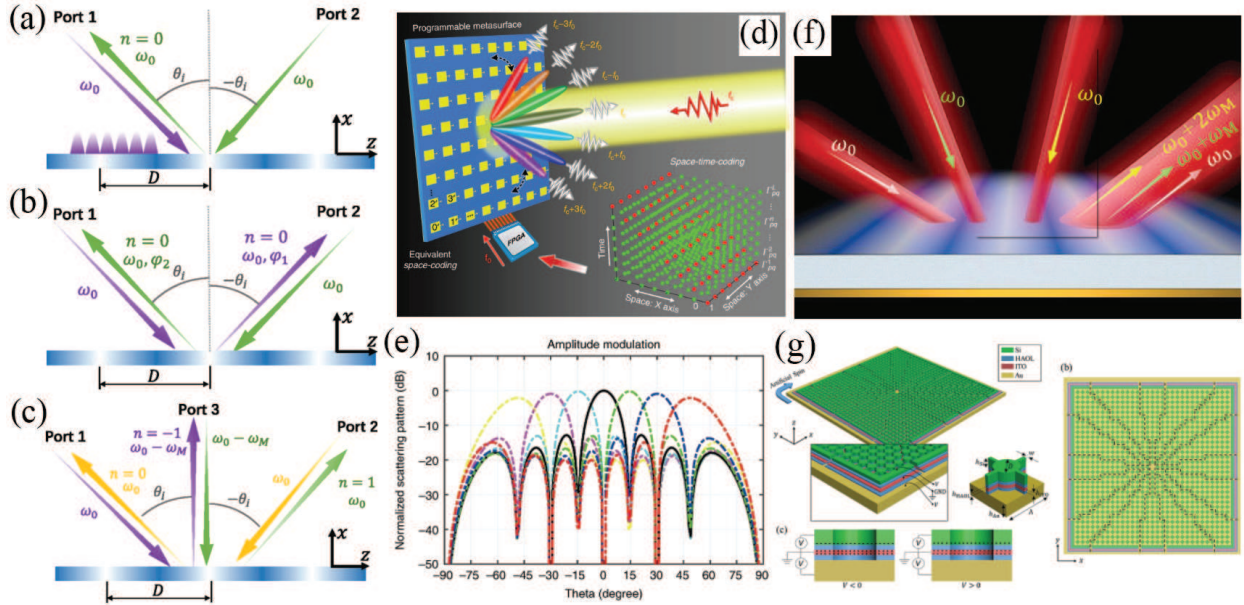}
    \caption{Examples of applications of space-time metasurfaces: (a) An isolator based on the unidirectional excitation of evanescent modes; (b) Waves scattering off a STM experience a nonreciprocal phase shift; (c) A STM-based circulator~\cite{wang2020theory}. (d) Illustration of an experimentally realized space-time coding metasurface, consisting of a square array of individual voltage-controlled elements with binary impedance values, enabling encoding and imprinting of arbitrary phase and amplitude modulation onto different scattered harmonics for (e) beam-steering and shaping~\cite{zhang2018space}. (f) Illustration of wave power combining using STMs: incoming waves with frequency $\omega_0$ coming from a discrete range of different angles can be scattered towards the same outgoing direction by imprinting the necessary frequency-wavevector shifts via the STM. (g) A proposal of an angular momentum-biased metasurface for generation of orbital angular momentum states: each azimuthal section is temporally modulated with a different relative phase, imparting the necessary angular-momentum bias. Figures adapted from Refs.~\citeonline{wang2020theory,zhang2018space,barati2020topological}}
    \label{fig:spacetime_metasurfaces}
\end{figure}

Systems such as isolators [Fig.~\ref{fig:spacetime_metasurfaces}(a)], nonreciprocal phase shifters [Fig.~\ref{fig:spacetime_metasurfaces}(b)] and circulators [Fig.~\ref{fig:spacetime_metasurfaces}(c)] using space-time modulated metasurfaces were designed, extending magnet-free nonreciprocity to surface implementations~\cite{hadad2015space,hadad2016breaking,shaltoutTimevarying2015,wang2020theory}. A promising idea to exploit these concepts in practice, at least at low frequencies, has come in the form of space-time-coding digital metasurfaces [Fig.~\ref{fig:spacetime_metasurfaces}(d)]: these systems consist of arrays of elements whose impedances are designed to take a discrete set of values, so that the overall response of the structure can be encoded as an array of bits (if only two values exist for each element)~\cite{zhang2019breaking}. Switching the individual impedances of a voltage-controlled array of such elements can enable exquisite control over the metasurfaces response in a system whose design can be systematically described using information theory concepts to tailor shape and direction of electromagnetic beams [Fig.~\ref{fig:spacetime_metasurfaces}(e)]~\cite{zhang2018space}, as well as to perform photonic analog computing~\cite{rajabalipanah2021analog}. Furthermore, close to the luminal regime, opportunities for nonreciprocal hyperbolic propagation~\cite{mazor2019one}, as well as vacuum \v{C}erenkov radiation~\cite{oue2021v} using space-time metasurfaces have recently been highlighted. 

Another application of space-time metasurfaces that was recently proposed is that of power combining of waves, a long-standing challenge in laser technology. Here the strategy consists of using a spatiotemporal modulation of a surface to effectively trade the difference in angle of incidence between a discrete set of incoming waves with identical frequency in exchange for a difference between frequencies of the outgoing waves, which emerge, at the same angle, and at an equally spaced set of frequencies, as illustrated in Fig.~\ref{fig:spacetime_metasurfaces}(f)~\cite{wang2021space}. More exotic ideas include angular-momentum-biased metasurfaces [Fig.~\ref{fig:spacetime_metasurfaces}(g)], which can impart a geometric phase on the impinging waves, and engineer photonic states with finite orbital angular momentum and optical vortices~\cite{barati2020topological,sedeh2020time}. Finally, the concept of space-time metasurfaces has recently been imported into the quantum realm, promising full control of spatial, temporal and spin degrees of freedom for non-classical light, with several opportunities for the generation of entangled photonic states, thereby anticipating a whole new scenery for this field to expand towards~\cite{kortcamp2021space}.

\subsection{Further directions}

Among new directions for space-time media, dispersive effects offer additional knobs for mode-engineering~\cite{chamanaraUnusual2018}, for which much still remains to be explored. Interestingly, spatial dispersion may be effectively engineered into a material via time-modulation, just as temporal dispersion arises from spatial structure~\cite{torrent2020strong}. Chiral versions of synthetically moving and amplifying electromagnetic media, mimicking the Archimedean screw for fluids, have recently been proposed, and may be realized with circularly polarized pump-probe experiments~\cite{galiffi2021archimedes}. Topology also occupies a prime seat, with Floquet topological insulators having been realized in acoustics~\cite{fleuryFloquet2016}, and theoretically proposed for light~\cite{he2019floquet}. Further related opportunities have been demonstrated for photonic realizations of the Aharonov-Bohm effect~\cite{fangPhotonic2012}, as well as higher dimensional topological effects such as Weyl points (topological invariants for 3D structures) in spatially 2D systems, realized by exploiting a synthetic frequency dimension~\cite{linPhotonic2016}. The idea of space-time media has also been formalized in the interesting mathematical formulation of ``field patterns". These objects arise from the consecutive spatial and temporal scattering in a material that features a checkerboard-like structure of its response parameters in space and time, and may offer new angles to study space-time metamaterials in the future~\cite{milton2017field,mattei2017field,LurieBook}. Finally, beside photonic systems, non-electromagnetic dynamical metamaterials have already acquired significant momentum, with several theoretical proposals~\cite{torrentNonreciprocal2018,nassarQuantization2018,liTransfer2019} and experimental realizations, including asymmetric charge diffusion~\cite{camachoAchieving2020}, nonreciprocity~\cite{fleurySound2014,wangObservation2018,chen2021efficient}, Floquet topological insulators~\cite{fleuryFloquet2016} and topological pumping~\cite{xuPhysical2020} amongst many others. In particular, digitally controlled acoustic meta-atoms have recently attracted significant interest to probe time-modulation physics in acoustics~\cite{cho2020digitally,wen2020asymmetric}.

Having discussed the wealth of potential new physics enabled by time-varying media, we now turn our attention to the latest technological advances towards experimental implementations in photonics, and their related challenges. 


\section{Experimental advances and challenges in all-optical implementations}

\label{sec:experiments}

The experimental realisation of time-varying effects constitutes a new, broad and fruitful field of research across the wider wave-physics and engineering spectrum. Implementation in optics, and more particularly in nanoscale architectures and metasurfaces offers new opportunities to address both fundamental open questions in wave physics as well as material science, and potentially groundbreaking technological pathways, such as low-footprint ultrafast optical modulators and nonreciprocal components. This is particularly challenging due to the need for modulating a medium on a temporal scale similar to the optical frequency of the light field. Time-varying effects have been demonstrated in mechanical~\cite{Bacot2019ProNatAcSciPhase}, magnetic~\cite{Schultheiss2021PRLTime}, acousto-optic~\cite{Kittlaus2018NatPhotNonreciprocal}, opto-mechanical~\cite{shenExperimental2016} and electronic systems~\cite{liraElectrically2012}. In particular, electromagnetic metasurfaces can be modulated via mechanical actuation, chemical reactions, or phase-change materials, as well as electrically~\cite{Shaltout2019ScienceSpatiotemporal}. Nevertheless, their slow modulation speed compared to near-optical timescales makes these poor candidates for achieving sizable frequency shifts and other time-varying effects in the visible or near-infrared ranges. The following section focuses on experimental realisations of time-varying media in these optical frequency ranges. Emphasis is placed on metasurfaces as these have subwavelength thicknesses that allow for purely temporal modulations of the medium. In a metasurface, the various constraints of bulk media such as loss or self-broadening are avoided, and the propagation of the modulating or probe beam in the medium does not need to be taken into account. We will review in a first step the state of the art of time-varying photonics experiments using nonlinear optical processes in semiconductors, more specifically their most common implementations: epsilon near zero (ENZ) materials, hybrid ENZ platforms and high-index dielectrics. We will then discuss emerging paths to implement all-optical ultrafast modulation within new structures or materials such as quantum wells, magnetic materials, multilayered metamaterials and 2D materials.


\subsection{Photocarrier excitation and nonlinear optical modulation} 

Thanks to their wide variety of implementations and frameworks, as well as their ultrafast nature, nonlinear optical interactions have proven to be a fertile ground for the development of time-varying systems at optical frequencies. In particular, modulation effects originated by short laser pulses interacting with a nonlinear medium occur on sub-ps timescales, well beyond the reach of electro-optic modulation, and very close to the time of an optical cycle (1-10~fs regime). Optical modulation of a medium stems from a change in $\varepsilon$, the relative permittivity,  due to the nonlinear response in a material. In semiconductors, these effects are often driven by out-of-equilibrium electronic populations, called hot-electrons. Nonlinear effects in semiconductors arise because of electron excitation of either real or imaginary states. Transitions through real states are realized by photocarrier excitation, that is intraband or interband transitions of the electrons within the active medium. The redistribution of electrons within the valence and/or conduction bands will affect the permittivity and thus the refractive index of the material on timescales limited either by the rate of transfer of energy from the light beam to the electrons or by the light pulse duration. This allows for sub-ps modulation of $\varepsilon$ in a variety of media of interest for photonics, including dielectrics and metals. Transitions through virtual states are on the other hand much faster, but lack the strength of photocarrier excitations: nonlinear processes via virtual states, such as four-wave-mixing and sum-frequency-generation, usually exhibit low efficiencies. Both photocarrier excitation and nonlinear optical modulation induce ultrafast modulation of $\varepsilon$, generally limited by the pulse duration and not material-limited. In the next section we will discuss recent progress in nonlinear optical modulation for time-varying effects in two main platforms: epsilon-near-zero media and high-index dielectrics.


   \subsubsection{Epsilon-Near-Zero materials, transparent conducting oxides and nonlinear optics}

\begin{figure}[H]
    \centering
    \includegraphics[width=\textwidth]{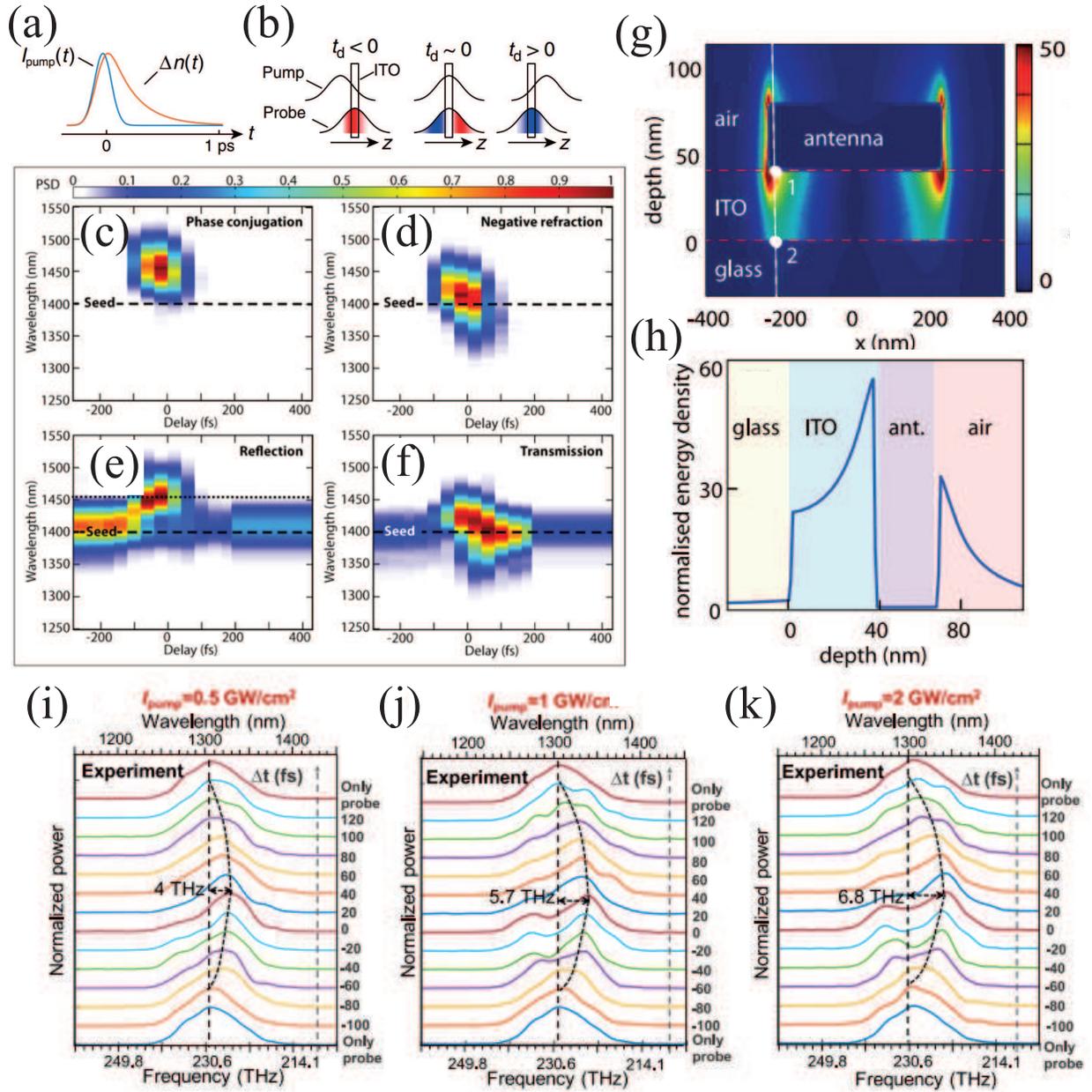}
    \caption{\textbf{(a,b)} Concept of photocarrier excitation and time-refraction in ITO~\cite{Zhou2020NatCommBroadband}: (a) A pump pulse transfers energy to the electrons in the conduction band, leading to a temporal variation in the refractive index. (b) As the probe experiences different indices at different delays, its spectral content will be shifted (redshift for negative delay, blueshift for positive). \textbf{(c-f)} Measured time-refraction, phase conjugation and negative refraction signals from a 500 nm thick AZO slab pumped by 105 fs pulses, at an energy of 770 \power in a degenerate pump-probe experiment at 1400 nm~\cite{Bruno2020AppSciBroad}. \textbf{(g)} Strong coupling between a plasmonic antenna and an ENZ thin film~\cite{Bruno2020PRLNegative}: electric field $|E|^2$ distribution obtained from FDTD for an ITO layer thickness of 40 nm and Au antenna length of 400 nm. \textbf{(h)} Resulting field distribution as a function of metasurface depth. \textbf{(i-k)} Probe spectrum at various delays for a central wavelength of 1304 nm for 50 fs pulses in a strongly-coupled plasmonic antenna-ENZ system~\cite{Pang2021NanoLettAdiabatic}. The increase in modulating pump intensities (0.5, 1 and 2 \power) leads to a shift near zero delay. Figures adapted from Refs. \citeonline{Zhou2020NatCommBroadband,Bruno2020AppSciBroad,Bruno2020PRLNegative,Pang2021NanoLettAdiabatic}.}
    \label{fig:expFig1}
\end{figure}

ENZ materials exhibit various interesting properties around their ENZ frequency, where the real part of their permittivity crosses zero, while the imaginary part stays relatively low. ENZ media feature guided and plasmonic modes for field enhancement, slow-light effects in waveguides and phase-matching relaxation~\cite{Kinsey2019NatRevMatNear}. Of particular interest are transparent conducting oxides (TCOs), degenerately doped semiconductors whose band structures allow for ENZ frequencies in the near-IR with strong optical nonlinearities. Indium Tin Oxide (ITO) and Aluminium Zinc Oxide (AZO) are well known, silicon-technology compatible TCOs with very strong nonlinear optical response~\cite{alamLarge2016,Caspani2016PRLEnhanced,Luk2015APLEnhanced,Carnemolla2018OptMatExpDegenerate} which makes them good candidates as platforms for time-varying metasurfaces. Particularly, intraband transitions in ITO lead to a strong redistribution of the effective mass of conducting electrons and thus an efficient shifting of the plasma frequency in the material's Drude dispersion as shown in Fig.~\ref{fig:expFig1}(a). As shown by Alam et al. in 2016~\cite{alamLarge2016}, order of unity modulation of the refractive index can be achieved in ITO. Keeping in mind that for an instantaneous change of index, the frequency shift due to time-refraction is proportional to $\Delta n/n$, where $n$ is the refractive index and $\Delta n$ the index change, strong frequency shifts can be achieved in ITO due to its low refractive index near its ENZ frequency, even for small changes $\Delta n$.

Time-refraction in pump probe experiments has been shown in bulk ITO and AZO thin-films~\cite{Ferrera2018JofOptUltrafast,Vezzoli2018PRLOptical,Bruno2020AppSciBroad,Zhou2020NatCommBroadband,Bohn2021Spatiotemporal}, with frequency shifts as strong as 58 THz~\cite{Bruno2020AppSciBroad} when the probe field and the modulation overlap [see diagram in Fig.~\ref{fig:expFig1}(b)]. Time-refraction not only applies to the reflected and transmitted probe pulse, but also to its phase conjugation and negative refraction, as shown in Fig.~\ref{fig:expFig1}(c-f). In the works of Ferrera et al.~\cite{Ferrera2018JofOptUltrafast}, Vezzoli et al.~\cite{Vezzoli2018PRLOptical} and Bruno et al.~\cite{Bruno2020AppSciBroad}, negatively refracted and phase-conjugated signals were recorded simultaneously with the time-refracted signal, with the internal efficiencies going above unity for these ultrafast, purely time-varying signals~\cite{Vezzoli2018PRLOptical}. This demonstrated the potential of these TCOs for time-varying applications, yet the strength of the modulation and signal was mostly achieved thanks to a significant propagation within the medium, and was hampered by losses in the medium (for ITO, the penetration depth of light around the ENZ frequency is about 100 nm). Time-refraction in a 80 nm thick ITO metasurface was demonstrated in Liu et al.'s work~\cite{Liu2021ACSPhotPhoton} in 2021, which in turn highlighted the need for a new understanding of the role of saturation of photocarrier excitation in the modulation of permittivity as large field intensities are now confined to significantly smaller portions of space. In this work it was shown that, as expected, a shorter pulse will lead to a stronger shift in frequency due to a larger $\wrt n/\wrt t$ during the short propagation time within the ITO slab, but more interestingly that saturation happens at lower energies for longer pulses. This indicates that the ultimate material response time for short and intense modulation constitutes still an open research question.

Nanostructuring ENZ materials is a powerful strategy  to increase the optical field enhancement and lead to efficient ultrafast modulation as demonstrated by Guo et al.'s work~\cite{Guo2016NatCommLarge} with ITO nanorods. In this work, the thickness of the medium is 2.6 $\mu$m, a length sufficient to cause significant loss upon propagation. An alternative to nanostructuring the ITO to increase the field enhancement consists of exploiting the plasmonic mode exhibited by flat ENZ thin films~\cite{Vassant2012OptExpBerreman}. In Bohn et al.'s work~\cite{Bohn2021NatCommAll}, the plasmonic ENZ mode was excited using a Kretschmann configuration, leading to a change in reflectance of 45\%.

While ENZ materials allow for large nonlinear responses and time-modulation, their large refractive index mismatch with air or larger-than-one-index media calls for advanced photonic architectures to ensure efficient coupling as we will address in the next section.


    \subsubsection{Hybrid platforms for time-varying experiments}

To compensate for the lack of interaction volume in a metasurface, good in-coupling and enhancement of the electric field is necessary to achieve strong nonlinear optical modulation. ENZ material-only metasurfaces lack strong resonant behaviour, and this has prompted the community to include additional elements to achieve efficient time modulation. Well-understood and easy-to-realize, plasmonic-ENZ metasurfaces provide light coupling and local field enhancement to otherwise impedance-mismatched ENZ materials.

Plasmonic nanoantennas couple efficiently propagating radiation to their near-field.
When plasmonic resonances are excited,  electromagnetic hot-spots form in the surrounding media. For this reason, metallic and more particularly Au nanoantennas have been used to couple light from the far-field to a thin ENZ substrate, where the strong field enhancement leads to efficient nonlinear modulation of the medium. In addition, strong coupling of localized plasmon resonances in Au antennas with the plasmonic modes of an ENZ film can be achieved~\cite{Bruno2020PRLNegative} (see Fig.~\ref{fig:expFig1}(g-h)). The coupled antenna-ENZ thin film enhances the field in the ENZ medium, and strong frequency shifts originating from time-refraction, as well as efficient negative refraction, have been measured from ITO films~\cite{Bruno2020PRLNegative,alamLarge2018,Liu2021ACSPhotPhoton} with a record 11.2 THz frequency shift being recorded at a comparatively low power of 4 \power in Pang et al.'s work~\cite{Pang2021NanoLettAdiabatic} [Fig.~\ref{fig:expFig1}(i-k)]. Such time-varying metasurfaces are at the moment limited by the low damage threshold of the plasmonic antennas, the damage threshold of ENZ materials such as ITO or AZO being much higher. 

In order to circumvent damage threshold constraints, architectures where the optical field is localized in the ENZ layer only have been explored: Au films can be used as a perfectly conducting layer in the near-IR and IR, beneath the ENZ thin film to increase its coupling to free space. Even though this is an impedance-matching effect, it can be understood as a superposition of the suppression of reflection for p-polarized light at the Brewster angle and the suppression of transmission from the reflective layer. Such a system was used by Yang et al.~\cite{Yang2019NatPhysHigh}: a layer of In:CdO with an ENZ frequency at 2.1 $\mu$m exhibited large nonlinearities when illuminated at its Brewster angle, with the 9th harmonic being generated and measured from the metasurface. Time-varying effects are noticeable in the harmonic spectrum, with for example the 5th harmonic exhibiting two peaks: a small one at $5f$ and a large, time-refracted one 48 THz below arising from the modulation of the index by the pump and the shifted frequencies being upconverted (that is a shift of about 2 THz of the probe signal due to time-refraction).

Hybrid plasmonic-ENZ systems thus present an improvement in realising time-varying metasurfaces using the favorable properties of ENZ materials and other TCOs. This calls for further investigation of the implementation of ENZ properties in other types of metasurfaces, such as the multilayered systems presented in section \ref{sec:multilayered}.



\begin{figure}[H]
    \centering
    \includegraphics[width=\linewidth]{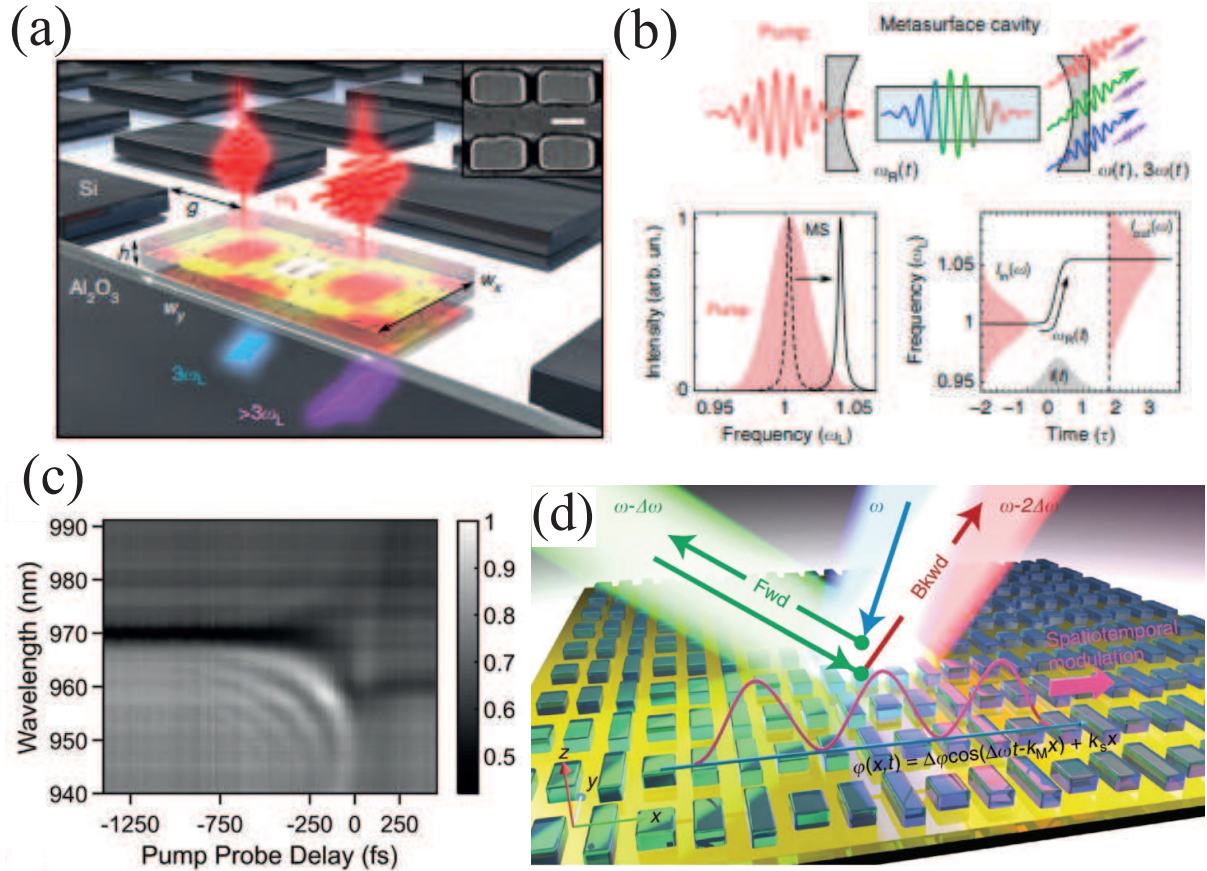}
    \caption{\textbf{(a)} Schematic of an Si metasurface engineered to support collective high-Q Fano resonances~\cite{Shcherbakov2019NatCommPhoton}. \textbf{(b)} Schematic of the self-induced blue shift of harmonics in the resulting Si nanoantenna cavity: photons in the antenna will undergo a blueshift due to the rapidly changing permittivity of the medium before being upconverted via third-harmonic generation. \textbf{(c)} Measured spectrum evolution for a 80 fs pump pulse length in a GaAs high-$Q$ metasurface~\cite{Karl2020NanoLettFrequency}. The fringes are measured on only one side of the spectrum due to the breaking of time-reversal asymmetry caused by the modulation. (d) Schematic of a nonreciprocal Si metasurface~\cite{Guo2019LightSciAppNonreciprocal}: the travelling wave modulation breaks the space-time symmetry of the reflection phase change. Figures adapted from Refs. \citeonline{Shcherbakov2019NatCommPhoton,Karl2020NanoLettFrequency,Guo2019LightSciAppNonreciprocal}}
    \label{fig:expFig2}
\end{figure}

        
    \subsubsection{High-index dielectric metasurfaces as time-varying media}

In the quest for resonant-enhanced materials for time-modulation, high-index dielectrics provide an alternative to ENZ structures, as they offer possibilities for various resonant architectures and a high damage threshold. More in particular, nanoantennas with Mie resonances or bound states in the continuum allow for strong time-modulation by combining efficient coupling to the active medium and a good field enhancement. That is because even though $\Delta n/n$ is low in a high-index dielectric, the change of phase of the metasurface can be strong when the material is modulated near a resonance. Shifts in the resonance of a photonic crystal~\cite{Husko2009APLUltrafast} were demonstrated, as well as in Mie-resonant systems, in a range of materials ranging from Si~\cite{Shcherbakov2015NanoLettUltrafast,Makarov2015NanoLettTuning,DellaValle2017ACSPhotNonlinear} to GaAs~\cite{Shcherbakov2017NatCommUltrafast}, GaP~\cite{Grinblat2020SciAdvEfficient} and Ge~\cite{Lemasters2021AdvOptMatDeep}. GaAs and Si exhibited stronger modulation efficiencies as they feature a direct band gap in the visible to near-IR region, while Si relies on intraband transitions, less efficient for this class of materials~\cite{Shcherbakov2017NatCommUltrafast}. 

Time-refraction was observed both in Si~\cite{Shcherbakov2019NatCommPhoton} and GaAs~\cite{Karl2020NanoLettFrequency} high-$Q$ nanoantenna metasurfaces. Shcherbakov et al.~\cite{Shcherbakov2019NatCommPhoton} reported a 8.3 THz shift in the third harmonic signal from a Si metasurface with a collective Fano resonance as shown in Fig.~\ref{fig:expFig2}(a), i.e. around 2 THz shift in the self-modulated pump signal. As depicted in Fig.~\ref{fig:expFig2}(b) this is explained by the upconversion to third harmonic of the time-refracted pump field, with the change of index of the nanoantenna cavity originating from the pump itself as well. This is a smaller shift than the best results achieved with combined plasmonic antenna-ENZ systems, but comparable to what was achieved in bulk ITO~\cite{Liu2021ACSPhotPhoton} and In:CdO on Au~\cite{Yang2019NatPhysHigh}. It is also worth noting that the system operates at 11 \power, underlining the higher damage threshold of dielectric antennas in comparison to plasmonic antennas.

Though high-index dielectric antennas exhibit higher damage threshold and provide more flexibility thanks to the tunability of the resonances by nanostructuring, as opposed to bulk ENZ properties, the interplay between the width of the resonance and that of the modulating pulse must be considered carefully. The higher the $Q$ factor, the stronger the potential for a modulation as the field enhancement is higher at resonance, yet if the duration of the modulating optical pulse is made shorter than the resonance lifetime, in order to accelerate the modulation and obtain stronger time-varying effects, its spectral content will exceed the resonance bandwidth and only a modest portion will interact. This calls for nondegenerate pump-probe experiments to show the full-potential of time-varying media: Karl et al.~\cite{Karl2020NanoLettFrequency} recorded the clear appearance of fringes in the probe spectrum due to time-varying effects as shown in Fig.~\ref{fig:expFig2}(c), by independently controlling the pump pulse length via pulse chirping and the spectral content of the probe via spectral filtering of a supercontinuum pulse. 

One can also take advantage of the phase-shift induced by a high-index nanoantenna array: Guo et al.~\cite{Guo2019LightSciAppNonreciprocal} demonstrated nonreciprocal light reflection in a Si nanobar metasurface. To this end, a spacetime modulation was induced by the interference and beating of two pump beams with a 6 nm difference in central wavelength and a nonreciprocal frequency shift was measured from the reflection of the forward and backward-propagating probe [Fig.~\ref{fig:expFig2}(d)].

Thanks to their strong, tunable resonances, high-index dielectric metasurfaces have proven to be a reliable platform for time-varying experiments. Although third-order nonlinearities in these materials are weaker than in ITO or AZO, and nanostructuring puts a lower cap on damage threshold and interaction volume, these metasurfaces bring concurrence to ENZ media thanks to the maturity of nanoantenna fabrication technology and the control of the scattering phase enabled by such systems.


    
\subsection{New leads for time-varying metasurfaces}

It is clear that better and more efficient material platforms for time-varying media are sought after and will likely appear in the near future, to allow for more efficient modulations, larger bandwidth, and higher damage thresholds. In this final section we identify a few promising candidates, namely quantum well polaritons, magneto-optical modulation, multilayered ENZ metamaterials and 2D materials, exhibiting the potential to enhance time-modulation effects and we discuss their respective strengths and weaknesses.


    \subsubsection{Quantum well polaritons}
    
Mann et al.~\cite{Mann2021OpticaUltrafast} showed in 2021 the pulse-limited modulation of intersubband polaritonic metasurfaces, using the coupling between patch antennas and intersubband transitions in multi-quantum wells (MQWs). The antenna resonance matches the transition dipole moment of the MQWs, which leads to Rabi splitting at low intensities. As the intensity increases, the ground state is depleted, which induces a change in coupling and absorption properties. The system operates at intensities between 70 kW/cm$^2$ and 700 kW/cm$^2$, with a change in absorption of order of unity at the antenna resonant frequency. Though the change in reflection is quite weak at such powers (at best 8\% here), it is worth noting that the intensities here are much lower than those used in other nonlinear optical experiments for time-modulation. On the other hand, the recovery time is dictated by the relaxation time of the excited state of the MQWs, here of 1.7 ps. In comparison, the modulation recovery time in unsaturated ITO is 360 fs~\cite{Alam2016ScienceLarge}.


    \subsubsection{Magneto-optical modulation} 
    
Though the mutual effects of electric and magnetic fields are quite weak, one can consider using the magneto-optical Kerr effect (MOKE) or Faraday effect (MOFE) to achieve ultrafast time-dynamics and thus time-varying physics in a magnetic medium. MOKE (MOFE) consists in the rotation of reflected (transmitted) light by a magnetic field~\cite{Maccaferri2020JofAppPhyNanoscale,Kirilyuk2010RevModPhysUltrafast}. A first stone was laid when Beaurepaire et al.~\cite{Beaurepaire1996PRLUltrafast} showed in 1996 sub-ps switching of spin in ferromagnets using a MOKE configuration. Ultrafast spin switching was demonstrated without magnetic fields using a circularly-polarized optical pulse~\cite{Kimel2005NatureUltrafast}, exploiting the dynamics of the inverse Faraday effect. Particularly, Stanciu et al.~\cite{Stanciu2007PRLAll} demonstrated a 40 fs all-optical switching of magnetization. Though in this experiment the origin of the switching was finally found out to be originating from the heating of the medium by the laser pulses, Mangin et al.~\cite{Mangin2014NatMatEngineered} later showed helicity-dependent switching in various magnetic media, independently of the threshold switching temperature of the medium. This opened a new path towards ultrafast magnetic memory writing and other applications of optical switching of magnets.
    
Experiments involving the use of a magnetic field use a common figure of merit, the $\delta$ parameter, defined as the percentage change of reflection when the magnetisation is reversed. That is for a magnetic field $\textbf{M}$ and a reflection $R$, $\delta=[R(\textbf{M})-R(-\textbf{M})]/R(\textbf{0})$. This value is usually quite low in bulk ferromagnetic thin films, ranging from $10^{-5}$ to $10^{-3}$. Nanostructuring can help increase this figure of merit: the coupling of surface plasmon polaritons from an Au grating to a ferromagnetic substrate allows for a stronger MOKE effect~\cite{Belotelov2011NatNanoEnhanced}, leading to a $\delta$ parameter of up to 24 \%~\cite{Belotelov2013NatCommPlasmon}. In other works, plasmonic antennas in a magnetic field were used to control the transmission of chiral light~\cite{Zubritskaya2017NanoLettMagnetic}, and dielectric nanoantennas were engineered via their electric and magnetic Mie resonances to achieve large Faraday rotation~\cite{Christofi2018OptLettGiant}. Yet, these time-modulations of the medium linked to the value of $\delta$ do require a switching of the magnetisation, which happens on a much slower scale (100 ps to ns~\cite{Zubritskaya2017NanoLettMagnetic, Landau1935PZSOn}) than the optical modulation from the MOKE itself. Hence, the ultrafast aspect of the modulation may only come from the rotation of the polarisation of light at a given, fixed magnetisation.
    
In a parallel stream of research, ultrafast spin currents have been excited in magnetized thin films and heterostructures using optical pulses, paving the way for THz emitter technologies~\cite{Kampfrath2010NatPhotCoherent,Kampfrath2013NatNanoTerahertz}. Photocurrents were generated on a timescale of 330 fs in Huisman et al.'s work~\cite{Huisman2016NatNanoFemtosecond}. As the magnetic field does not require switching and can be maintained constant, the modulation can be considered as entirely optical and ultrafast. Qiu et al. demonstrated similar spin current generation in an antiferromagnetic slab in the absence of magnetic field~\cite{Qiu2020NatPhysUltrafast}, thanks to second-order optical nonlinearities. For a more detailed review of THz emitters and wave generation the authors recommend the review by Feng et al.~\cite{Feng2021JofAppPhySpintronic}. While the potential of magneto-optical effects for ultrafast phenomena has been demonstrated, their translation into the framework of time-varying media remains unexplored, which calls for a new push in investigations and experiments in this direction.

    
    \subsubsection{Multilayered metamaterials}
    \label{sec:multilayered}
    
Multilayered metal-dielectric structures are of great interest thanks to their tunability to different regimes via fabrication, such as ENZ or hyperbolic. In addition, their simple architecture for fabrication is well-understood from effective-medium theory~\cite{Orlov2014PhotNanoControlling}. Artificial ENZ metamaterials exhibiting Dirac cone-like dispersion~\cite{Moitra2013NatPhotRealization} or multilayered thin films make valuable candidates for time-varying experiments, as they exhibit the favorable ENZ physics, are tunable and have the potential to exploit other meta-properties, e.g. good coupling to far-field radiation. Sub-ps modulation of the effective permittivity of a multilayered Au/TiO\textsubscript{2} metamaterial was demonstrated by Rashed et al.~\cite{Rashed2020PRBHot}. In parallel, enhancement of nonlinear properties around the ENZ frequency of a multilayered Ag/SiO\textsubscript{2}~\cite{Suresh2020ACSPhotEnhanced}, as well as ultrafast modulation of absorption~\cite{Acharyya2021AdvMatIntUltrafast} were reported. This enhancement is explained by the dependence of the effective third-order nonlinear susceptibility on the inverse of the medium's effective index, which reaches a minimum in the vicinity of the multilayered structure's ENZ frequency. In a similar spirit, ultrafast modulation of a metal-insulator-metal nanocavity was observed around the low-energy ENZ point~\cite{Kuttruff2020CommPhyUltrafast}, but it was found that the switching speed was limited by the carrier density and heat capacity of the metal. The authors suggested the use of TCO-dielectric structures to accelerate the electron dynamics of the process and achieve faster modulation. Alternatively, multilayered ENZ metamaterials can be used to pin down and control the plasmonic resonance of coupled antennas depending on the ENZ frequency~\cite{Habib2020NanoControlling}, with optical modulation allowing for control of the antenna resonance via the multilayered substrate. 

In a different paradigm, multilayered metal-dielectric antennas have been engineered to support ultra-small mode volumes~\cite{Indukuri2019ASCNanoUltrasmall}, and in this way couple light efficiently to a WS\textsubscript{2} monolayer~\cite{Indukuri2020ASCAppNanoMattWS2}. WS\textsubscript{2} and 2D materials are a promising class of media for time-varying experiments as we discuss in the following section. Although monolayers feature properties favorable to modulation, their intrinsic atomic-scale thickness limits their coupling to light: hyperbolic metamaterials such as a multilayered metal-dielectric antenna can overcome this barrier and boost optical modulation. Furthermore, efficient second and third harmonic generation was measured from multilayered Au/SiO\textsubscript{2} nanoantennas~\cite{Maccaferri2021ACSPhotEnhanced}. The efficiency of the second harmonic was proved to originate from the multiple metal-dielectric interfaces rather than symmetry-breaking, allowing for a polarisation-independent implementation of multilayered antennas.

To sum up, multilayered ENZ metamaterials constitute promising candidates for time-varying experiments as they exhibit ultrafast modulation and strong nonlinear properties, while offering options for nanostructuring and resonance engineering.




    \subsubsection{2D materials}
    \label{sec:2dmaterials}
    
A modern topic in nanophotonics, 2D materials exhibit many interesting properties including high refractive index and ease of nanostructuring~\cite{Verre2019NatNanoTransition}, unconventional band structure and carrier dynamics such as Dirac cones~\cite{Castro2009RevModPhysThe}, and excitonic physics as well as ultra-high carrier mobility~\cite{Schaibley2016NatRevMatValleytronics}, resulting in strong optical nonlinearities useful for material modulation~\cite{You2018NanoNonlinear,Yu2017AdvMat2D}. Transition metal dichalcogenides (TMDs), black phosphorus (BP) and graphene feature sub-ps carrier excitation times~\cite{Poellmann2015NatMatResonant,Pogna2016ACSNanoPhoto,Wang2016ACSNanoUltrafast} in single layer structures. In particular, TMDs offer excitonic resonances with short recombination times, while monolayer BP exhibits a tunable direct band gap from the visible to the mid-IR. Additionally, BP's anisotropic response has also been shown to undergo modulation under laser excitation~\cite{Ge2015NanolettDynamical}. 
TMDs exhibit a variety of useful properties for optical modulation~\cite{Autere2018AdvMatNonlinear}, with monolayer and low-dimensional TMDs exhibiting stronger nonlinear properties than their bulk counterparts. In Nie et al.'s work~\cite{Nie2014ACSNanoUltrafast}, a 10 fs pulse of about 1 \power was shown to excite photocarriers in 20 fs in low-dimensional MoS\textsubscript{2}, though the change in transmission was only of 0.8\%. Large second and third-order nonlinearities have also been measured in low dimensional MoS\textsubscript{2}~\cite{Kumar2013PRBSecond,Saynatjoki2017NatCommUltra}, MoSe\textsubscript{2}~\cite{Le2016AdPNonlinear}, WS\textsubscript{2}~\cite{Janisch2014SciRepExtraordinary,Torres20162DMatThird} and WSe\textsubscript{2}~\cite{Ribeiro_20152DMatSecond}, as well as strong tunability of the refractive index in WSe\textsubscript{2}, MoSe\textsubscript{2} and MoS\textsubscript{2}~\cite{Wang2015PhotResTunable}. Additionally, the strong nonlinear properties in MoSe\textsubscript{2} can be exploited thanks to the comparatively high damage threshold of the monolayer as demonstrated by Tam et al.~\cite{Le2016AdPNonlinear}. Ultimately, applications are limited by the low-dimensionality of these materials which leads to small interaction volume and short interaction time. On the other hand, nanostructuring can again be used to enhance the response from these systems: enhanced light-matter interaction has been shown in monolayers coupled with plasmonic nanoantennas~\cite{Kern2015ACSPhotNanoantenna,Spreyer2020NanoSecond,Chen2019AppMatToResonance} as well as with a Si waveguide~\cite{Chen2017LightSciAppEnhanced}. Nonlinear antennas built from bulk TMDs could also provide a platform for time-varying experiments in the same fashion as GaAs or Si~\cite{Busschaert2020ACSPhotTransition}. Another alternative would be van der Waals heterostructures, stacked layers of TMDs which have also exhibited short rise times of about 50 fs~\cite{Hong2014NatNanoUltrafast,Ceballos2014ACSNanoUltrafast}, with no dependence of the response time on the twisting angle between layers~\cite{Wang2016ACSNanoInterlayer,Zhu2017NanoLettInterfacial}.More information on the specific spin and valley dynamics at the origin of the ultrafast dynamics in van der Vaals heterostructures can be found in Jin et al.'s review~\cite{Jin2018NatNanoUltrafast}.

\subsection{Outlook}

All-optical implementations of time-varying metasurfaces are of great interest for the creation of new miniaturized technologies for the control of light both in space and time. ENZ materials such as ITO and AZO, along with high-index dielectrics, have proven to be solid platforms for time-varying experiments and will surely be the ground for more complex investigations. These nanophotonic architectures can boost both light coupling to the nonlinear medium and field enhancement. Multilayered ENZ metamaterials could provide new solutions to the practical problems posed by the nature of more classical time-varying systems, such as resonance-engineering or high permittivity contrast. New materials and systems could pave the way for further development of time-varying experiments, expanding the spectral range of operation, increasing the efficiency and the damage threshold. Quantum well polaritons and monolayer 2D materials could exhibit time-varying effects at lower energies, while magneto-optical effects would open a new framework including time-varying magnetic effects, which are ignored in current nonlinear optical experiments. 


\section{Conclusions}

In this Review Article, we have presented a comprehensive overview of photonic time-varying media. We started by reviewing the basic phenomenological and mathematical considerations rooted in the behaviour of Maxwell's Equations in the presence of temporal material discontinuities, discussing the main directions of ongoing research on electromagnetic time-switching for several applications such as time-reversal, energy manipulation, frequency conversion, bandwidth enhancement and wave routing, amongst many others. We then continued our discussion to consider photonic time-crystals, discussing the basic phenomenology underlying periodic time-scattering and parametric amplification to develop an insight which we deployed in addressing more advanced instances of time-modulation in particular for topological physics, non-Hermitian systems and disorder, concluding with some brief remarks on time-modulated surfaces. We then extended our discussion to combinations of spatial and temporal degrees of freedom, providing an overview of the basic properties of space-time crystals, and their application to nonreciprocity, as well as the engineering of synthetic motion and its applications to optical drag and giant bianisotropic responses, highlighting the concept of luminal amplification and spatiotemporal localization as a new form of wave amplification physically distinct from the conventional parametric gain, and concluding with an overview of the wealth of opportunities and applications for space-time metasurfaces. Finally, we reviewed some of the most successful materials and paradigms for experimental realizations of time-varying effects in the visible and IR, indicating some of the most promising avenues recently unveiled for future optical experiments with time-varying media. While the latter undoubtedly constitute the greatest long-term challenge for this rising field of research, we believe that, as in many other instances of scientific enquiry, the quest for time-varying photonic systems will prove a prolific one from both the point of view of unveiling fundamentally novel wave phenomena and for revealing new and unexpected windows of opportunity for technological advancement. 

\subsection* {Acknowledgments}
E.G. acknowledges funding from the Engineering and Physical Sciences Research Council via an EPSRC Doctoral Prize Fellowship (Grant No.
EP/T51780X/1) and a Junior Fellowship of the Simons
Society of Fellows (855344,EG).
P.A.H. and M. S. acknowledge funding from Funda\c c\~ao para a Ci\^encia e a Tecnologia and Instituto de Telecomunica\c c\~oes under project UIDB/50008/2020 and the CEEC Individual program from Funda\c c\~ao para a Ci\^encia e a Tecnologia with reference CEECIND/02947/2020. 
R.S. J.P. and S.V. acknowledges funding from the Engineering and Physical Sciences Research Council (EP/V048880). J.P. acknowledges funding from the Gordon and Betty More Foundation. A.A., S.Y. and H.L. acknowledge funding from the Department of Defense, the Simons Foundation and the Air Force Office of Scientific Research MURI program.


\bibliography{bibliography}   
\bibliographystyle{spiejour}   




\end{spacing}
\end{document}